\documentclass[useAMS,usenatbib]{mn2e}

\usepackage{graphicx}
\graphicspath{{/Documents/LaTeX/figs/}}
\usepackage{subfigure}
\usepackage{caption}
\newcommand {\aplt} {\ {\raise-.5ex\hbox{$\buildrel<\over\sim$}}\ }
\usepackage[abs]{overpic}

 \title[Galactic Interaction as the Trigger for the Young Radio Galaxy MRC
B1221$-$423]{Galactic Interaction as the Trigger for the Young Radio
Galaxy MRC B1221$-$423
\thanks{Based on observations made with ESO Telescopes
 at the La Silla Observatory, program 084.B-0677(A)}
}
 \author[C.\ S.\ Anderson, H.\ M.\ Johnston, R.\ W.\ Hunstead]
{C.\ S.\ Anderson$^{1}$\thanks{E-mail to:
c.anderson@physics.usyd.edu.au}, 
H.\ M.\ Johnston$^{1}$,
R.\ W.\ Hunstead$^{1}$\\
 $^{1}$Sydney Institute for Astronomy, School of Physics, University of 
Sydney, NSW 2006, Australia}

\begin{document}

\date{date accepted date received here}

\pagerange{\pageref{firstpage}--\pageref{lastpage}} \pubyear{2002}

\maketitle

\label{firstpage}

\begin{abstract}
\label{abstract}
 Mergers between a massive galaxy and a small gas-rich companion (minor
mergers) have been proposed as a viable mechanism for triggering radio
emission in an active galaxy. Until now the problem has been catching this
sequence of events as they occur. With MRC B1221$-$423 we have an active
radio galaxy that has only recently been triggered, and a companion galaxy
that provides the ``smoking gun". Using spectroscopic data taken with the
VIMOS Integral Field Unit detector on the European Southern Observatory's
Very Large Telescope, we have examined the distribution, ionization state
and kinematics of ionized gas in this interacting system. We have also
modelled the stellar continuum with synthesised spectra of stellar
populations of different ages. From our study of the ionized gas, we have
derived preliminary models for the geometry of the interaction, analysed
the kinematic behaviour of the ionized gas, and examined the ionization
mechanisms at work throughout the system. Our modelling of the stellar
continuum allowed us to identify and date distinct stellar populations
within the galaxy pair. We find evidence of multiple episodes of
widespread starburst activity, and by dating these populations, we provide
tentative insight into the history of the interaction.

\end{abstract}


\section{Introduction}
\label{Intro}
 One of the most dramatic but poorly understood stages in the evolution of
galaxies is the formation of an Active Galactic Nucleus (AGN; henceforth
used for both the singular and plural). This energetic phenomenon is
thought to be powered by material accreting onto a supermassive black hole
(SMBH) lying at the core of the host galaxy. It is now becoming clear that
SMBHs and the galaxies they inhabit share an intimate co-evolutionary
history, suggesting that the presence of an AGN affects the host galaxies
in important and unexpected ways.  Examples include modulation of star
formation via radio-mode feedback (e.g. Silk and Rees 1998, Fabian 1999,
King 2003), the SMBH-to-bulge mass relation (Ferrarese \& Merritt 2000,
Gebhardt et al. 2000, Di Matteo et al. 2005), and acting to enrich the IGM
in metals (Germain et al. 2009, Barai et al. 2011), energetic particles
(Ginzburg \& Syrovatskii 1965, Hillas 1984), and magnetic fields (Jafelice
et al. 1992, Kronberg et al. 2001, and references therein). The processes
occurring in AGN are thus thought to play a fundamental role in driving
the evolution and ecology of galaxies. 

 Among the most significant gaps in our current understanding of AGN is
the triggering phase of the activity. A number of different mechanisms may
be responsible for AGN triggering, ranging from internal stochastic
processes in relatively low luminosity AGN (Ballantyne et al. 2006;
Hasinger 2008; Hopkins \& Hernquist 2009; Lutz et al. 2010) through to
full-scale galactic mergers in the most luminous examples (e.g. Barnes \&
Hernquist 1991; Mihos \& Hernquist 1996). To determine the relative
importance of environmental and internal processes will require spatially
resolved studies of large samples of galaxies hosting an AGN. Two upcoming
surveys will address this need: the CALIFA survey (S\'{a}nchez et al.\
2012) will perform integral field unit (IFU) observations of a
statistically well-defined sample of over 600 galaxies, while the SAMI
survey (Croom et al.\ 2012) will bring a powerful new combination of
multi-object and IFU spectrograph technology to bear on the study of over
5000 galaxies in a range of environments.

To date, arguably the most well studied and robust of the proposed
triggering mechanisms is the merger scenario. This model proposes that
gravitational interaction between two galaxies triggers a large amount of
gas to migrate to the core of the host galaxy, where it forms an accretion
disk around the SMBH in the galactic nucleus. The hot accretion disk and
the charged particles accelerated as a direct result of the accretion
process are then responsible for the prodigious power output that we
observe.

A number of lines of evidence provide support for the merger hypothesis.  
For example, AGN hosts have been found to often display disturbed
morphologies (Longair et al. 1995; Best et al. 1996, Koss et al. 2010,
Mezcua et al. 2012), and show evidence of significant bursts of ongoing or
recent global star formation (Veilleux 2001, and references therein;
Mezcua et al. 2012). These properties are generally interpreted as arising
from gravitational interaction with a neighbouring galaxy. In addition, a
number of statistical studies have pointed to a tendency for AGN hosts to
reside in over-dense regions or alongside companion galaxies of comparable
size (e.g. Koss et al. 2010). Though these observations would seem to
imply that galaxy mergers play an important role in AGN triggering, the
evidence is not completely clear-cut. For example, recent studies have
found no clear evidence for increased merger rates or an excess of nearby
companions for typical AGN hosts (e.g. Miller et al. 2003, Grogin et al.
2005, Li et al. 2006, Gabor et al. 2009). Recently, even the correlation
between a galaxy's AGN-hosting status and the level of morphological
disturbance it shows have been called into question by Cisternas et al.
(2011). Each of the studies above seem to indicate either zero or weak
association between AGN activity and the host galaxy's environment or
morphology. It is difficult to explain these results if the majority of
AGN are triggered through mergers between galaxies of comparable mass
(so-called `major mergers').

 In part to address the inconsistencies outlined above, it has been
suggested that so-called minor mergers (often defined as mergers between
galaxies with a mass ratio of \aplt1:3 --- see Stewart 2009; Lotz et al.\
2011) may play a significant role in triggering AGN activity (e.g.
Taniguchi 1999). This scenario proposes that an AGN can be triggered by
interactions between a large host galaxy and a dwarf companion, and it
directly addresses several of the objections to the major merger model
raised above. Minor mergers are known to be strongly associated with
kinematic disturbances that can trigger activity such as starbursts in
the host galaxy (e.g. L\'{o}pez-S\'{a}nchez 2010). It appears that most
galaxies are orbited by small companions (e.g. Zaritsky et al. 1997), and
computational modelling shows that minor mergers should be efficient at
driving gas into the core regions of the dominant galaxy in such scenarios
(Hernquist \& Mihos, 1995). However, the same characteristics that make
minor mergers desirable from a theoretical point of view make them
difficult to study observationally: Minor mergers produce intrinsically
less disturbance to the host galaxy making them hard to identify
morphologically, the systems will only appear disturbed for a shorter
period of time, and minor companions are low mass objects by definition,
making them intrinsically faint and more difficult to detect.

MRC B1221$-$423 is a system that has been identified as a massive galaxy
that is undergoing a minor merger with a low mass gas-rich companion
(Johnston et al. 2010). Furthermore, it contains a young compact steep
spectrum (CSS) source with an estimated age of $\sim10^5$ years (Safouris
et al. 2003). The youth of the CSS source means that this AGN has only
been triggered recently, and given that we now observe the system
undergoing a minor merger, it represents an extraordinary laboratory for
studying merger-induced AGN triggering. We obtained Integral Field Unit
(IFU) spectroscopy of the system using the VIsible Multi Object
Spectrograph (VIMOS) instrument on the Very Large Telescope (VLT) at the
European Southern Observatory (ESO), Chile. Using this dataset, we were
able to study the kinematics and ionization of the gas, as well as examine
the ages of the constituent stellar populations, to arrive at a tentative
model for the interaction.

Our spectroscopic study of the MRC B1221$-$423 system is presented as
follows. In section 2 we describe our observations and data reduction. In
section 3.1, we present results and analysis of the distribution and
kinematics of the ionized gas. In section 3.2, we describe the results
from modelling of the stellar continuum. In section 4, we discuss our
results in the light of previous work done on this system, and present our
conclusions. Throughout, we have assumed a flat $\Lambda$-CDM cosmology
with $H_0 = 71$\ km\ s$^{-1}\ $Mpc$^{-1}$, $\Omega_M = 0.27$ and
$\Omega_\Lambda = 0.73$. At the redshift of the galaxy, 1 arcsec
corresponds to a projected separation of 2.876 kpc.

\section{Observations and Data Reduction}
\subsection{Observations}
\label{Observations}
 Service mode observations of the MRC B1221$-$423 system were carried out
using the VIMOS instrument (Le F\`{e}vre et al. 2003), mounted on the `UT3
Melipal' telescope at the ESO VLT. The instrument was used in its IFU
spectrographic mode.

The observations were taken in six separate observation blocks (OBs), on
the nights of 2009-Dec-16, 2009-Dec-22 \& 2010-Jan-10. Each OB consisted
of two ÔobjectÕ exposures bracketing a single sky exposure. The object
exposures in each OB were jittered to ensure full coverage of the system
in the event of dead IFU fibres (hereafter, we will refer to single IFU
fibres as `spaxels'). Integration times for the object and sky exposures
were 1260 and 350 seconds respectively, giving a total on-object
integration time of 252 minutes.

\begin{figure*}\centering
\subfigure[]{\includegraphics[scale=0.25]{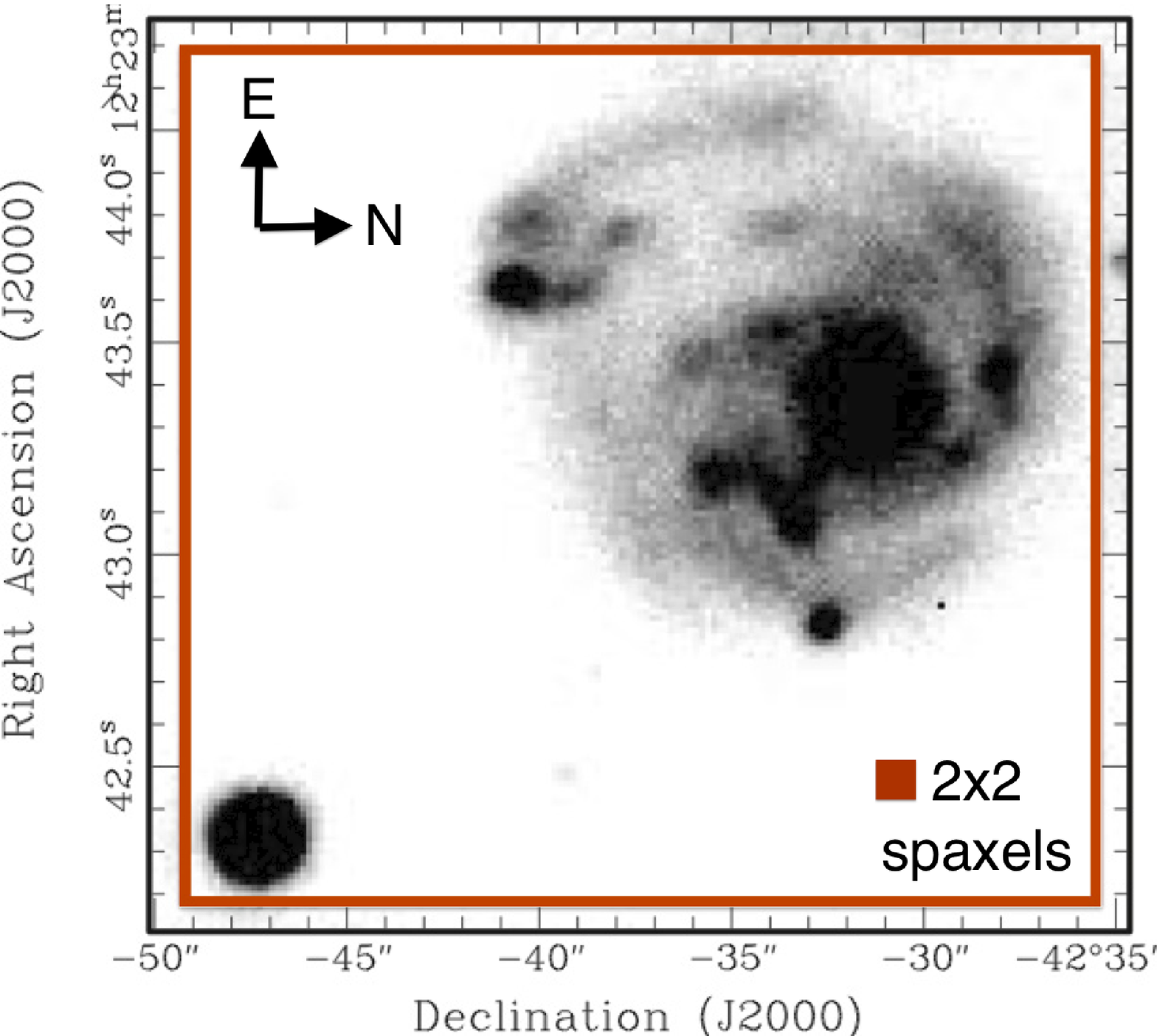}}
\hspace{0cm}
\subfigure[]{\includegraphics[scale=0.31]{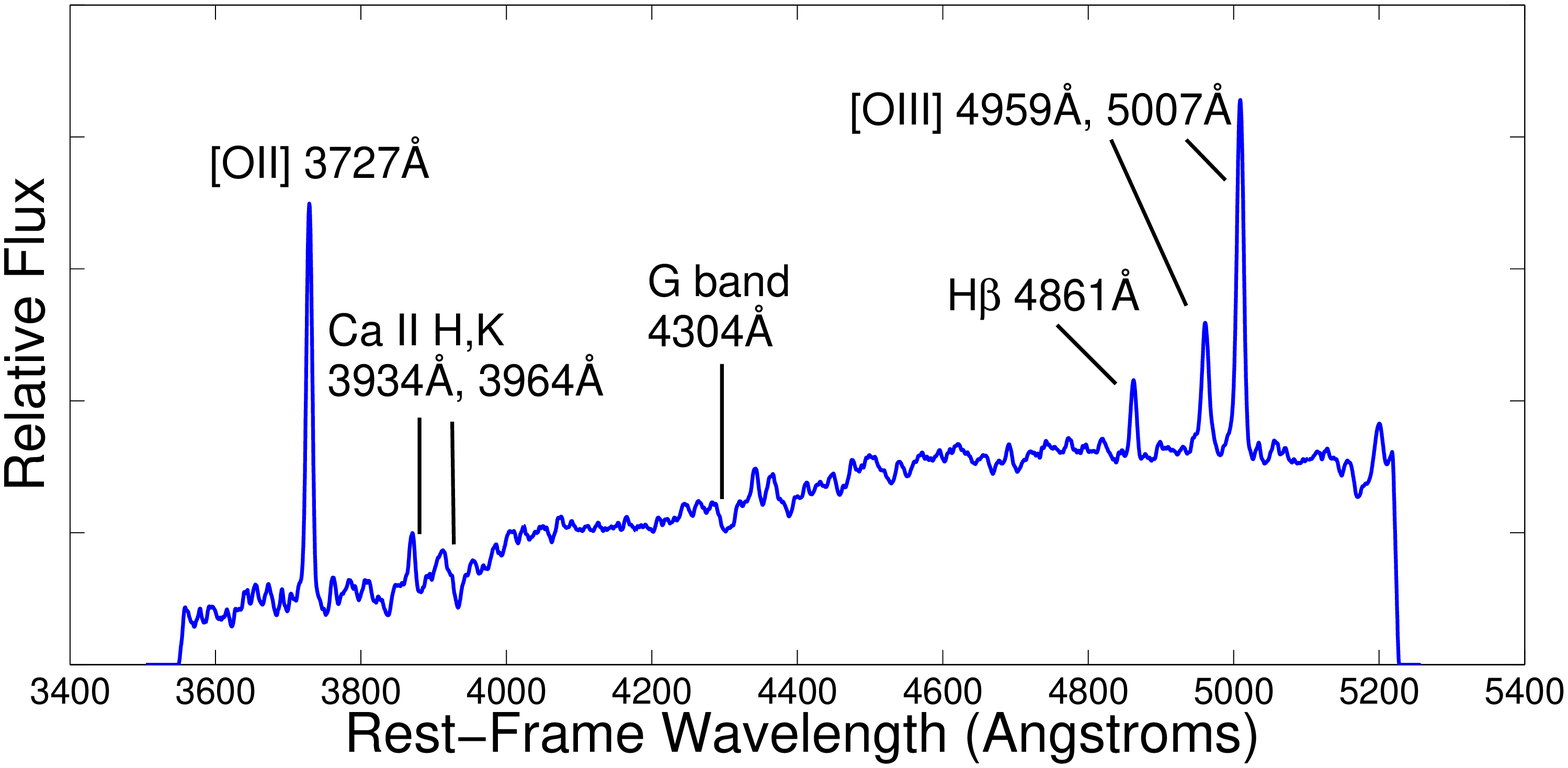}}
\hspace{0cm}
 \caption{\label{speccoverage}a) The spatial coverage of our VIMOS IFU
observations delimited by the red box; the spaxel scale is defined by a
solid red square. East is up and North is to the right. Image pixel scale
is that of the NTT optical image. The MRC B1221$-$423 system can be seen
in the North-East corner. The bright object in the South-West of the frame
is a Galactic star. b) Spectral coverage of our observations, showing a
spectrum of the nuclear region of the MRC B1221$-$423 system, with
wavelengths given in the rest frame of the system ($z = 0.1706$). The
spectrum is a composite of 20 spaxels covering the nuclear region. The
y-axis scale is in arbitrary units. }
 \end{figure*}

Our observations utilised the instrument in its low spatial resolution
mode coupled with the HR Blue grism. This configuration yields a
$40\times40$ spaxel IFU head for a total number of 1600 spaxels, with a
spatial resolution of $0.67''$ per spaxel and a spectral resolution of
2\,\AA\ at the central wavelength of the grism. The resulting field of
view (FOV) covered $27'' \times 27''$ on the sky
(Figure~\ref{speccoverage}a), while the spectral coverage was
4150-6200\AA\, corresponding to $\sim$3550-5220\AA\ in the rest frame of
the target galaxy (Figure~\ref{speccoverage}b). Data in each OB were taken
when the sky transparency was `clear', the seeing FWHM was $<1.2''$, the
moon illumination $<0.7$ (grey), and moon position $>30^{\circ}$ from the
target galaxy.

In addition, a number of calibration frames were taken contemporaneously
with the target observations. These included arc-lamp exposures, halogen
lamp flats, bias frames and observations of spectrophotometric standard
stars.

\subsection{Data Reduction}
\label{DataReduction}

The data reduction followed standard IFU reduction procedures as
implemented in the VIMOS Interactive Pipeline Graphical Interface (VIPGI),
a purpose-written program for reducing VIMOS data. The program itself and
the data reduction procedures are described in detail in both Scodeggio et
al. (2005) and the VIPGI User Manual. 
Briefly, the process involved bias
correction, flat-fielding and wavelength calibration of the IFU images
based on calibration data taken during observations. A spectrophotometric
calibration was then applied to correct for variation in spectral
sensitivity of the IFU system and obtain the absolute flux scale for
the data. The spectra were then collapsed to 1D, extracted, and corrected
for variation in the relative transmission of the IFU fibres. Finally, the
data for each OB were assembled into a datacube based on a mapping between
positions of fibres in the IFU head and spectra on the CCD.

We implemented the next step in our reduction process, the sky
subtraction, outside the VIPGI environment. Due to the extended nature of
our target, separate exposures of a nearby and relatively empty patch of
sky were obtained in addition to the object exposures. The sky subtraction
was applied to data in each OB separately, using the following procedure:

\begin{enumerate} 
 \item The calibrated data from each of the 1600 spectra forming the sky
exposure for a given OB were sorted according to the integrated flux and
noise levels over a defined region of the spectrum. The median 10\% were
retained while the rest were discarded, eliminating any sky spectra
recorded by fibres with poor transmission characteristics.
 \item The 160 retained spectra were averaged to form a `master' sky
spectrum.
 \item The master spectrum was individually scaled to each of the spaxels
in the two object exposures in each OB, based on the strength of the
prominent [O\,I]$\lambda$5577 sky emission line.
 \item The sky spectrum was subtracted from the object spectra.
\end{enumerate}

Data from individual OBs were aligned using their World Coordinate
System (WCS) header information and stacked to yield the final product
datacube. To assess the accuracy of the WCS alignment, we also manually
calculated the offsets between exposures based on the position of a bright
star located fortuitously in the FOV (Figure~\ref{speccoverage}a). The
accuracy of the alignment was then assessed for both methods by collapsing
the stacked datacube along the spectral axis and examining the 2D
autocorrelation function of the resulting image. The alignment based on
the WCS headers was found to be at least as accurate as the manual
determination.

 A correction for Galactic reddening was then applied to the data with
$A_V = 0.331$ mag. (Schlegel et al.\ 1998). The Calzetti et al.\ (2000)  
recipe was used for this purpose, being specifically formulated to
de-redden the spectra of starburst galaxies. The formulation states that
given an observed spectrum $f_{obs}(\lambda)$ and a reddening coefficient
$A_V$, we can calculate the de-reddened spectrum, $f_{int}(\lambda)$,
using the relations:

\[f_{int}(\lambda)=\frac{f_{obs}(\lambda)}{10^{-0.4A_{\lambda}}},\]
\noindent where
\[A_{\lambda}=\frac{A_V}{4.05 
}[2.659(-2.156+\frac{1.509}{\lambda}-\frac{0.198}{\lambda^2}+\frac{0.011}{\lambda^3})+4.05].\]

 An additional correction for reddening internal to the MRC B1221$-$423
system was made using the same formulation and estimates of $A_V$ made by
Johnston et al.\ (2005).  We applied a correction across the field based
on their estimated $A_V$ map, which ranged from close to 0 for many parts
of the system, to $\sim1$ for the knotty star-forming regions and the
companion, and up to 3 for the inner nucleus. We note here that in section
\ref{Fitting} where we model the stellar continuum, we allow $A_V$ to be a
semi-free parameter in the fit of intrinsic reddening to the uncorrected
data. We find that the values used for $A_V$ for our intrinsic reddening
correction described above for the various regions in the system come
close to providing the optimal fit between the stellar models and data.
This provides some level of {\it a posteriori\/} confirmation of the $A_V$
values we adopt from Johnston et al.\ (2005).

\section{Results}
\subsection{Ionized Gas}

Our observations allowed us to study the 2-dimensional distribution,
kinematics and ionization levels of the ionized gas across the MRC
B1221$-$423 system. To characterise the gaseous emission, Gaussian
profiles were fitted to prominent emission lines in the datacube in a
semi-automated fashion, using scripts developed for this purpose (provided
by R. Sharp, private communication, July 2010). This software uses an IDL
implementation of the AMOEBA downhill simplex fitting algorithm (Press et
al. 2007). Each line was fitted in isolation using a standard four
parameter Gaussian model:\\

$G=A_0 \exp(-x^2/2) + A_3$,\\

\noindent where $x= (\lambda-A_1)/A_2$, $A_0$ determines the amplitude of
the Gaussian peak, $A_1$ its central wavelength, $A_2$ its width, and
$A_3$ the continuum level. Suitable line and continuum fitting constraints
were defined in advance of fitting. Line fitting was assessed via a simple
comparison of the $\chi^2$ fits of the data with the emission-line model
and a constant continuum model. After fitting, 2-D line maps of the full
field were formed.

 Where strong discontinuities were observed, the fitted spectra were
examined visually to check for accuracy and whether fitted features were
genuine. In this way, fits were derived for the [O\,III]$\lambda$5007, 
[O\,III]$\lambda$4959, H$\beta$ and [O\,II]$\lambda$3727 emission lines for each of the 1600
spectra comprising the datacube.

Single Gaussian profiles were generally found to provide accurate fits to
the line profiles across most of the face of the system. The exception was
the high excitation [O\,III] doublet in the innermost $\sim$6\,kpc of the
nuclear regions. A routine capable of fitting two Gaussian components 
was found to provide a superior fit. However, detailed interpretation of
the gaseous kinematics that this implies is beyond the scope of this
paper.

With the fitting procedure completed, the fit parameters were recorded
along with associated spatial positions and spectral lines. These included
the peak amplitude, peak position, flux under the line at full-width
zero-intensity, line width at full-width half-maximum, and the continuum
flux density on either side of the line.

\subsubsection{Distribution of Ionized Gas}
\label{Distribution}

\begin{figure*}
\centering
\hspace{-0.4cm}
\vspace{0.2cm}
\subfigure[]{\includegraphics[scale=0.36]{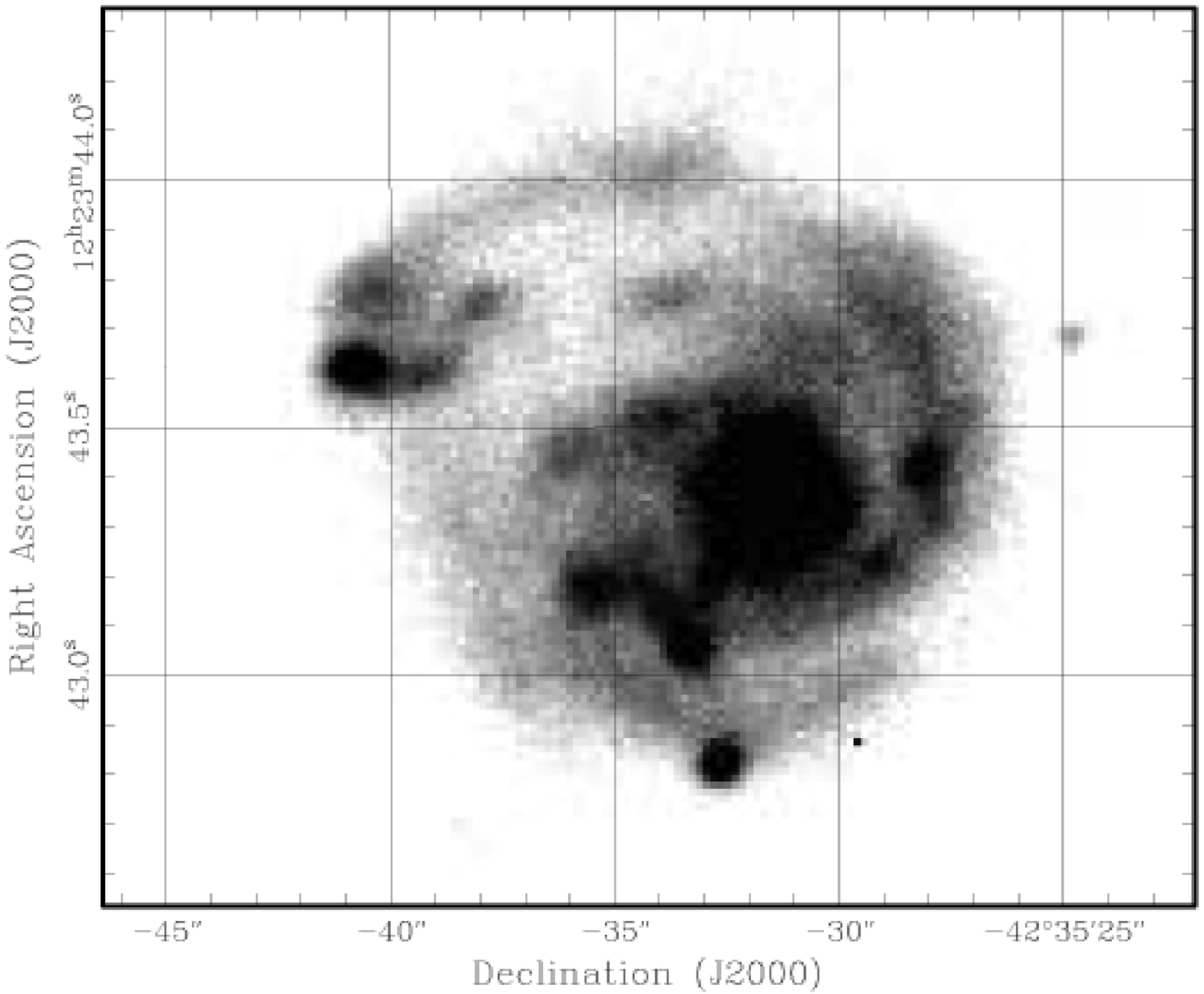}}
\hspace{0.7cm}
\subfigure[]{\includegraphics[scale=0.50]{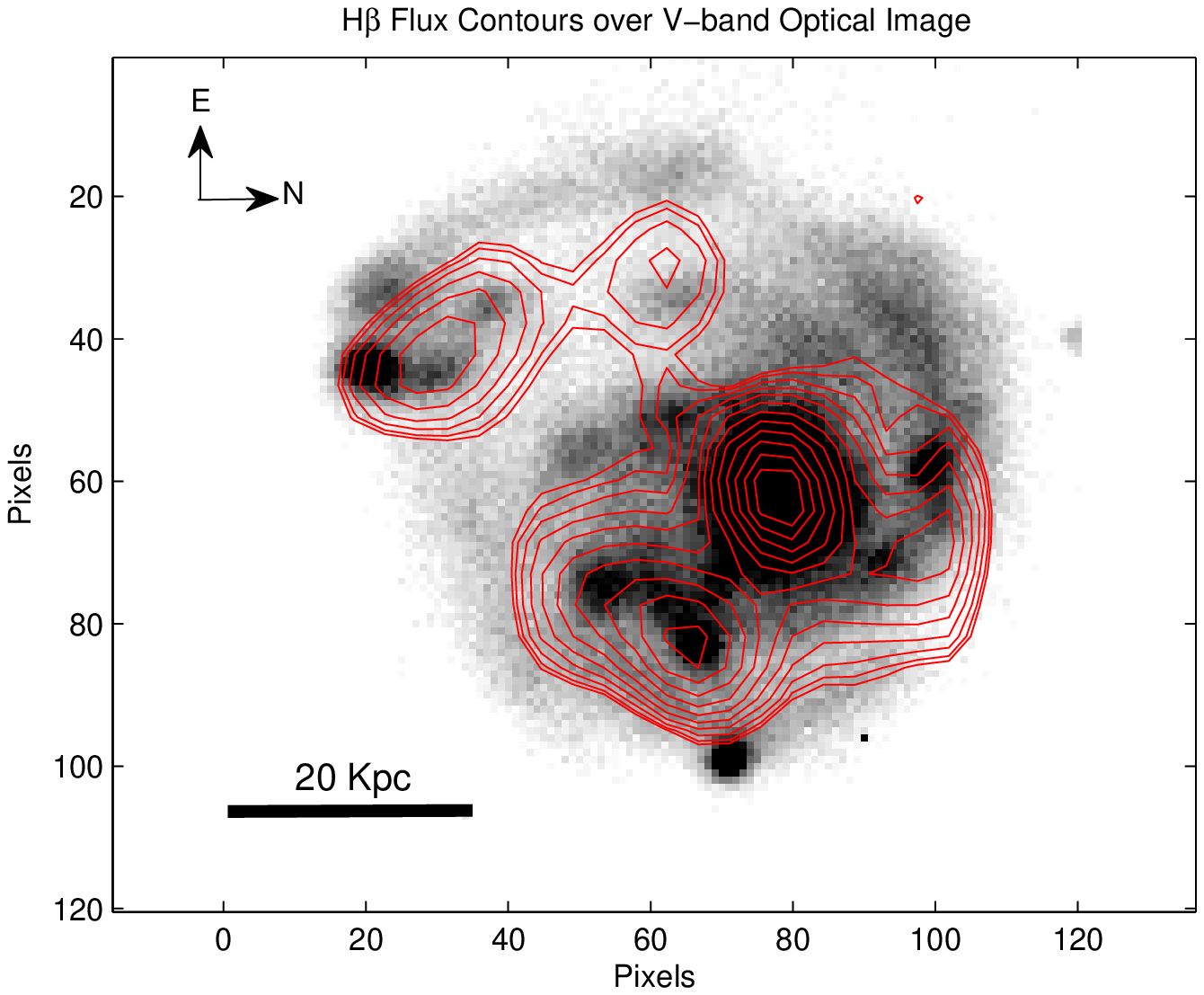}}
\hspace{0cm}
\subfigure[]{\includegraphics[scale=0.50]{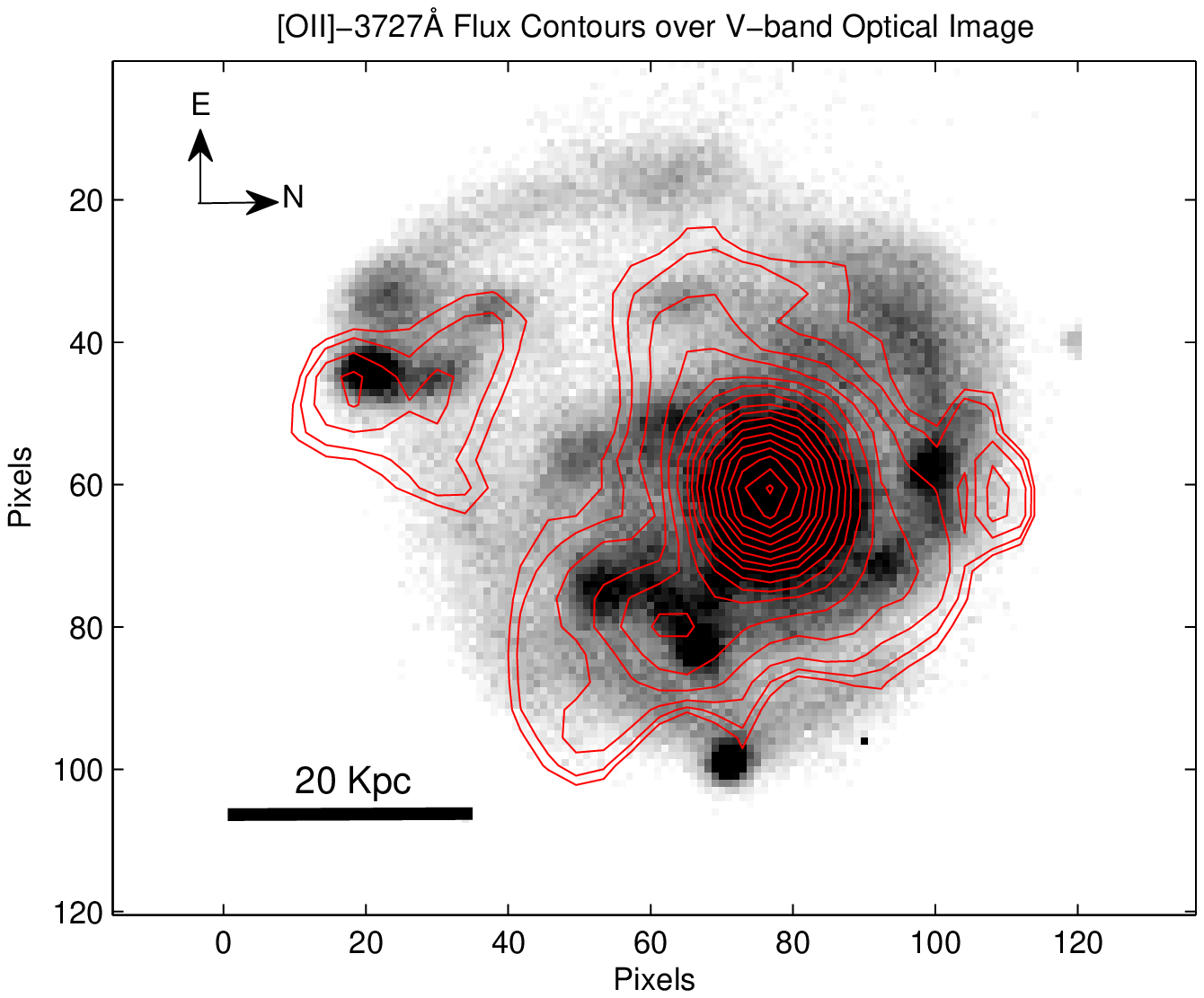}}
\hspace{0cm}
\subfigure[]{\includegraphics[scale=0.50]{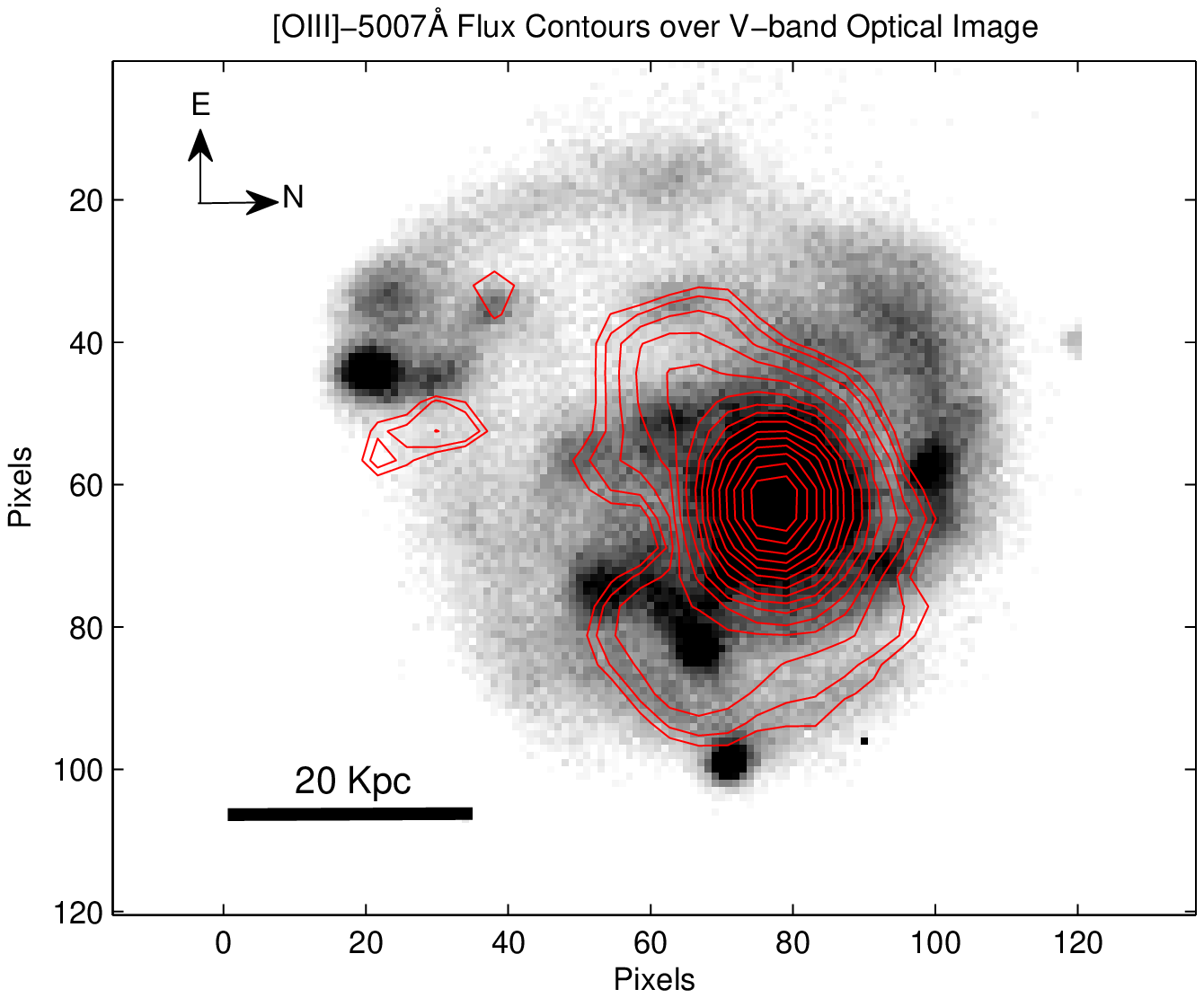}}
\hspace{0.75cm}
 \caption{\label{ContourFluxMaps} a) V-band NTT image of the MRC
B1221$-$423 system in greyscale.  The axis scales in the subsequent images
correspond to the pixel numbers in this NTT image, with overlaid
logarithmic contours, spaced by $\sqrt{2}$, as follows: b) H$\beta$
emission, over the interval 0.3 to 5.0 $\times 10^{-16}$\
erg\,s$^{-1}$\,cm$^{-2}$, c) [O\,II]$\lambda$3727 emission, over the
interval 1.0 to 36.0 $\times 10^{-16}$\ erg\,s$^{-1}$\,cm$^{-2}$, d)
[O\,III]$\lambda$5007 emission, over the interval 0.3 to 18.4 $\times
10^{-16}$\ erg\,s$^{-1}$\,cm$^{-2}$. All IFU data have been spatially
smoothed with a $3\times3$ spaxel boxcar.}
 \end{figure*}

Spatial maps of the emission intensity across the system were made for
each of the major emission lines in the spectral range.
Figure~\ref{ContourFluxMaps} shows the spectral line contours overlaid on
the V-band optical image.

The strongest overall emission feature in the spectral region observed is
the [O\,II]$\lambda$3727 line (Fig. 2c). The [O\,II] emission intensity
peaks strongly in the nuclear region of the galaxy, but significant
emission is seen throughout the system, most notably in the vicinity of
the nucleus, the bright knotty complexes to the west, and the companion
galaxy. Notwithstanding its limitations, [O\,II] emission is often used as
a tracer of star formation in galaxy systems, and the relative strength of
this emission line throughout much of the MRC B1221$-$423 system is
suggestive of widespread ongoing star formation.

H$\beta$ emission (Figure~\ref{ContourFluxMaps}b) is also found widely
throughout the system, and represents the strongest emission outside the
nucleus. The emission is strongly correlated with the bright knotty
complexes, which superficially take on the appearance of star-forming
regions embedded in spiral arms. It is well known, however, that the host
galaxies of CSS sources have absolute magnitudes consistent with
first-ranked ellipticals, and frequently show evidence of
interactions (eg. O'Dea 1998), with linear features that can mimic spiral
arms. Moreover, the H$\beta$ emission does not appear to match the optical
emission seen sweeping from the companion towards the host galaxy to the
east of the nucleus. The morphology and location of this feature is
suggestive of a tidal tail being pulled out as a consequence of the
gravitational interaction between host and companion. We speculate that
this tidal debris may wrap completely around the host galaxy, forming a
continuous shell of material with the knotty regions to the west of the
nucleus (see the toy model presented in section \ref{ModelGeometries}). If
so, the entire shell may represent gas and dust that has been fed into the
host after being tidally stripped from the companion. An alternative
possibility is that the shell may have been pulled out of the companion by
ram-pressure stripping as it passed through the extended regions of the
host on its orbit.

The fact that the H$\beta$ and V-band emission are not spatially
coincident in the tidal tail is notable, given that the H$\beta$ line is
typically associated with the presence of young stars. A speculative
interpretation is that the tidal tail of stars seen in the V-band image is
a relic of a previous orbit of the companion around the host, and is now
gas poor after star formation and gas stripping during the ensuing period.
The H$\beta$ emission would then trace the tidal tail pulled out by the
companion during the {\it current\/} orbit, and which, not having had time
to form stars in great numbers, is relatively dim in V-band emission. We
investigate this possibility further in section \ref{FittingResults}.

The [O\,III]$\lambda$5007 (Figure~\ref{ContourFluxMaps}d) emission shows
up primarily in the nucleus of the host galaxy. Given its ionization
energy of 35.12eV, we would expect [O\,III] to be largely confined to the
central regions of the host, where the hard UV continuum from the AGN is
sufficiently intense, or to regions surrounding massive stars in
star-forming regions. In addition to the nucleus, we detect significant
levels of emission from two extended regions to the South-East and
South-West of the nucleus. The region to the South-West may be correlated
with the knotty complexes in the V-band image, but the emission to the
South-East does not match any significant optical feature. This suggests
the presence of high energy processes not associated with star formation
occurring in this region; we return to this point in the following
section.

\subsubsection{Ionization Levels in the MRC B1221$-$423 System}
\label{ionization}

 Since our wavelength coverage did not include all the lines used in the
more commonly used emission-line diagnostics, we used a diagnostic diagram
and associated classification scheme devised by Lamareille (2010).
Briefly, the scheme can be used to differentiate between multiple AGN-like
and star-formation processes, by plotting log([O\,III]/H$\beta$) versus
log([O\,II]/H$\beta$). The classification regions are defined empirically,
based on the observational characteristics of a large sample of
extragalactic objects.
 
 A correction for intrinsic reddening in the MRC B1221$-$423 system was
first applied to the data, based on the Calzetti formulation (Calzetti et
al. 2000) and estimates for the reddening parameter $A_V$ derived by
Johnston et al. (2005).

\begin{figure*}
\centering
\subfigure[]{\includegraphics[scale=0.50]{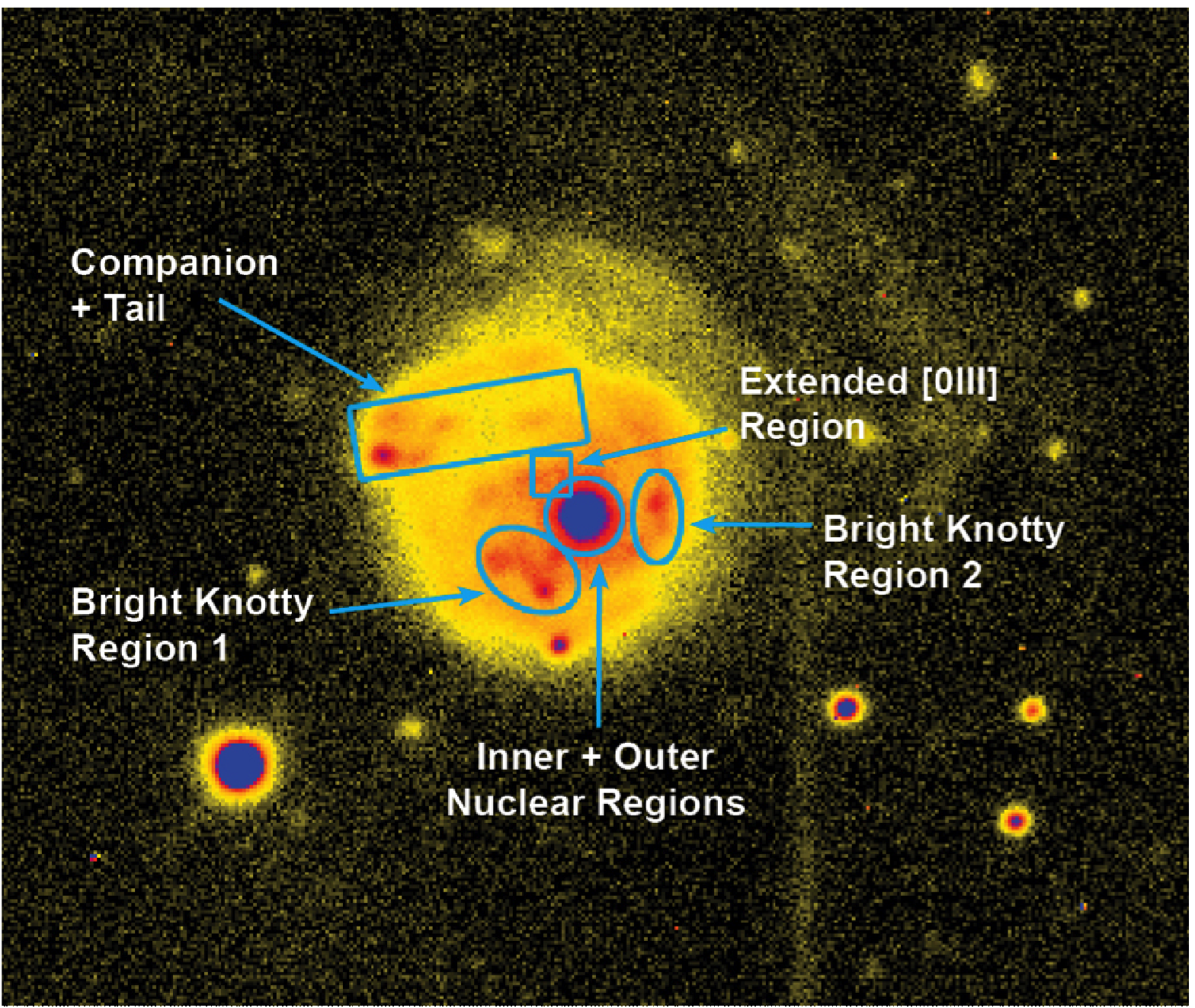}}
\hspace{0cm}
\subfigure[]{\includegraphics[scale=0.52]{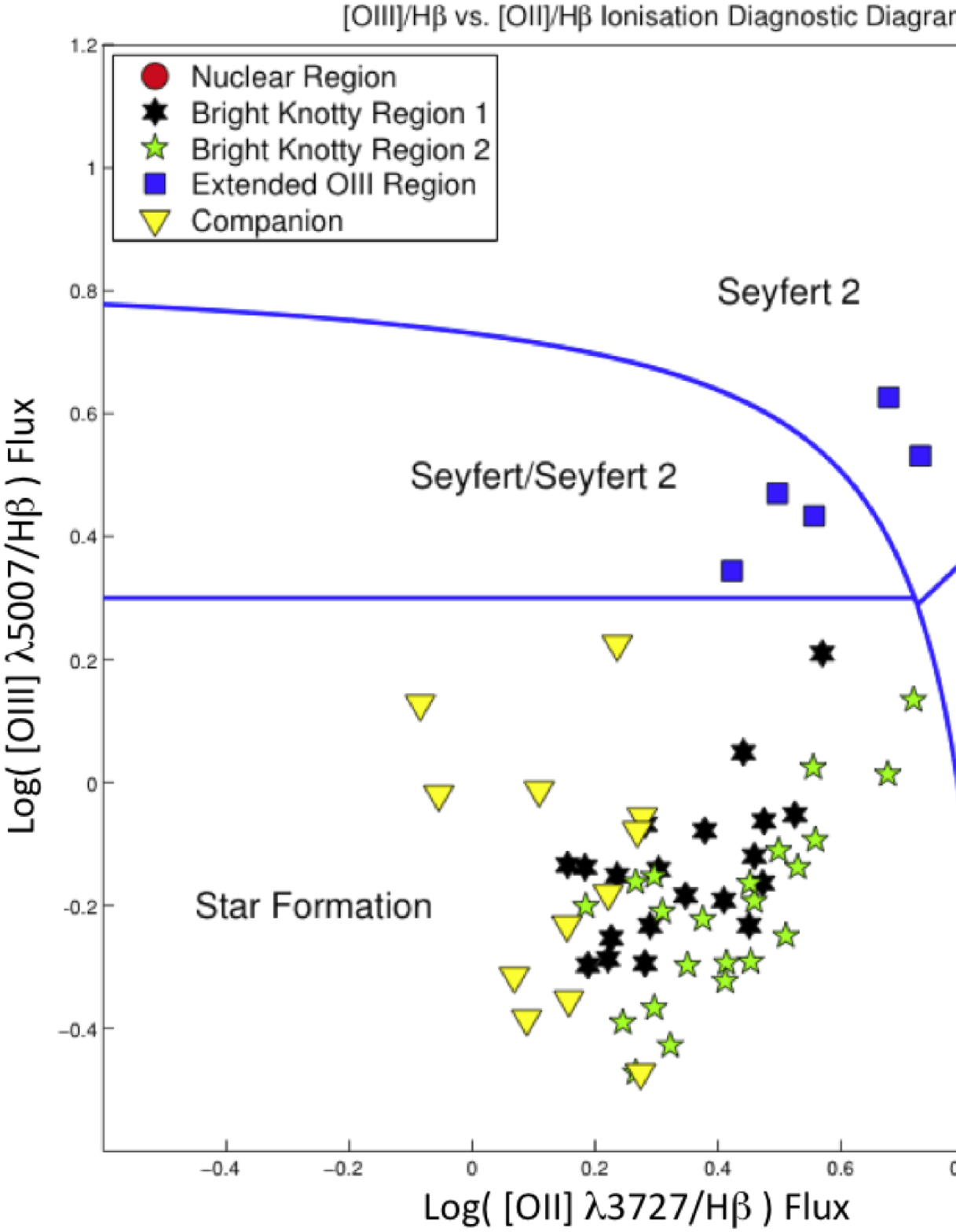}}
\hspace{0cm}

\caption{\label{Diagnostics} a) V-band image depicting the spatial regions
from which the spaxels were drawn for the analysis in section
\ref{ionization}. The false colour scheme was employed to maximise the
contrast of distinct morphological features. b) Lamareille diagnostic
diagram showing log([O\,III]/H$\beta$) versus log([O\,II]/H$\beta$) for
regions in which these values are defined. The blue lines demarcate the
diagnostic regions as defined by Lamareille (2010). The points separate
out cleanly into star-formation and AGN-like processes, showing strong
correlation between spatial location of the spaxel and the dominant
ionization process at work in the region.  }
 \end{figure*}

 Figure \ref{Diagnostics} shows the resulting diagram, with data points
colour-coded according to their spatial association with optical features
seen in V-band images (see inset, Figure~\ref{Diagnostics}). The data
points separate cleanly into several distinct regions across the diagram,
indicating that different ionization mechanisms are operating in different
spatial locations across the galaxy. The knotty regions have line ratios
consistent with those of star-forming galaxies, confirming the presence of
ongoing star formation in these regions. Likewise, line ratios indicative
of star formation are also found in the emission from the companion galaxy
and tidal tail. The hypothesis that this galactic system is undergoing a
gravitational interaction/merger is further supported by these results
which indicate the presence of widespread active star formation throughout
the system.

The nuclear region of the host galaxy maps to the LINER/Seyfert 2 regions
of the plot, consistent with the object's existing classification as a
LINER galaxy. A small region of extended [O\,III] emission to the
southeast of the nucleus (identified in section~\ref{Distribution}) maps
to the regions of the plot indicating AGN-type processes. This suggests
that higher energy processes operate in this region, potentially
associated with the AGN activity. Possibilities may include
hard-ultraviolet emission from the AGN accretion disk, shock-front
ionization from fast nuclear outflows or jets, or shock ionization from
gaseous inflows.

\subsection{Integrated Emission Line Fluxes}
\label{Integrated}

 In this section, we calculate the integrated flux for each of the three
main emission lines in our spectra, both in specific regions of the
interacting system and in the system as a whole.  From this we estimate
the global star-formation rates, H\,II mass and metallicity for those
regions in which the diagnostic emission-line ratios indicate pure star
formation and are not associated with AGN activity (see section
~\ref{ionization}).

The spatial regions adopted are identical to those used in our work on
stellar continuum fitting and are described in section~\ref{Continuum}
below. Spaxels were summed from the nuclear region of the galaxy as
defined in Fig.\ \ref{fig:BinRegionsa}(region 6), while the two knotty
regions, companion and extended [O\,III] region were all summed within the
limits of the regions depicted in Fig.\ \ref{fig:BinRegionsb} (regions 3
\& 4, 5 and 6 respectively).

 The integrated line + continuum fluxes in each spatial region were
obtained by summing the flux density over the relevant spectral regions
for each line. A continuum band adjacent to the lines and with the same
total spectral width, was then subtracted to yield the integrated emission
line fluxes.  Uncertainties were estimated from the noise in the continuum
bands, with appropriate scaling to take account of the subtraction
process. The H$\alpha$ luminosity was estimated by multiplying the
H$\beta$ flux by 2.8 (the theoretical case-B Balmer decrement), assuming a
luminosity distance of 770\,Mpc.

 We calculated the star formation rate (SFR) for each of the defined
regions based on the relationship between H$\alpha$ and SFR described by
Calzetti et al.\ (2007):

\[{\rm SFR (M_\odot\,yr^{-1})} = 5.3\times 10^{-42} 
L({\rm H}\alpha)_{corr}\,{\rm erg\,s^{-1}}.\]

\noindent We find a very high star-formation rate throughout the system,
especially in the knotty blue regions described earlier. The uncertainty
in the SFR quoted in Table\ \ref{intfluxtab} is based solely on the
uncertainty in the H$\beta$ flux and does not take account of the
significant uncertainty in the correction for intrinsic reddening in the
galaxy (section ~\ref{ionization}). Nevertheless, the SFR is undoubtedly
high, consistent with a gas-rich merger.

The H$\beta$ flux may also be used to calculate the H\,II mass of the
galaxy using the relationship described by P\'{e}rez-Montero (2002):

\[M{\rm _{H\,II} (M_\odot)} = 1.485\times 10^{-35} 
L({\rm H}\alpha)_{corr} \times(n_e/100),\]
 
\noindent where $n_e$(cm$^{-3}$) is the electron density. As expected for
a gas-rich merger, we find large reservoirs of ionised hydrogen,
particularly in the knotty regions. As shown in section ~\ref{kinematics}
below, this gas may be moving in large bulk flows through the system as a
result of the merger process.

 We include in Table\ \ref{intfluxtab} the gas-phase metallicities
calculated using the iterative method of Kobulnicky \& Kewley (2004); see
also L\'{o}pez-S\'{a}nchez \& Esteban (2010). Most of the B1221$-$423
system possesses very similar O/H metallicity, between 8.4--8.8 for all
regions examined assuming `lower branch' metallicities, or around 9.0 for
`upper branch' metallicities. Without additional information from the
[N\,II]/H$\alpha$ ratio, however, we are unable to determine which branch
our object lies on.


\begin{table*}
 \caption{\label{intfluxtab} Integrated fluxes and derived quantities for
selected regions of the MRC B1221$-$423 system (defined in section
~\ref{Continuum}), corrected for reddening.
 Listed uncertainties are based on the estimated errors in the
emission-line fits alone, and do not include uncertainties in the
reddening correction. The two values given for the O/H abundance 
correspond to the lower- and upper-branch values respectively; see text.  }
 \footnotesize
 \begin{minipage}[b]{1.0\textwidth}
 \centering
\begin{tabular}{lllllllllll} 
 \hline
 & & &\multicolumn{3}{c}{Flux ($10^{-16}$ erg\,s$^{-1}$\,cm$^{-2}$)} \\
 \cline{4-6}
 Region & Spaxels & $A_V$ &[O\,II] & H$\beta$ & [O\,III] & Lamareille & $L({\rm H}\beta$) & SFR & $M_{\rm H\,II}$ & 12+log(O/H)\\
  &  & mag. &$\lambda$3727 &  &$\lambda$5007 & (Fig. 3) & {\scriptsize $10^{40}$ erg\,s$^{-1}$} & {\scriptsize M$_\odot$/yr} & {\scriptsize $10^6$ 
  M$_\odot$} & \\
 \hline
Nucleus & 16 & 2.5 & $650 \pm 15$ & $49 \pm 3$ & $230 \pm 4$ & LINER/Sey & $34.9\pm1.9$ & --- & --- & --- \\
Knotty reg.1 & 24 & 1 & $71 \pm 13$ & $36 \pm 4$ & $14 \pm 2$ & SF & $25.3\pm2.4$ & $37.5\pm1.3$ & $70.8\pm2.4$ & 8.51, 9.07\\
Knotty reg.2 & 30 & 2 & $21 \pm 3$ & $8.9 \pm 0.7$ & $5.3 \pm 0.6$ & SF & $6.4\pm0.5$ & $9.5\pm0.3$ & $17.9\pm0.5$ & 8.79, 9.03 \\
Ext. [O\,III] & 9 & 1.5 & $13 \pm 4$ & $3.6 \pm 3.1$ & $12.3  \pm 1.0$ & Sey & $2.6\pm2.4$ & --- & --- & ---  \\
Companion & 56 & 1 & $10 \pm 3$ & $7.0 \pm 0.7$ & $7.6 \pm 0.6$ & SF & $5.0 \pm0.5$ & $7.4\pm0.3$ & $14.0\pm0.5$ &  8.44, 9.05 \\
Whole system & 396 & varies & $790 \pm 20$ & $117 \pm 6$ & $298 \pm 7$ & All  & $83\pm4$ & $> 54$ & $> 103$ & 8.58, 9.05  \\
 \hline
\end{tabular}
 \label{tab1}
 \end{minipage}
\end{table*}

\subsubsection{Global Kinematics of the Ionized Gas}
\label{kinematics}

The kinematics of the gas and stars provides an important test of the
merger hypothesis, yielding constraints on the possible geometry of the
interaction. Safouris et al.\ (2003) used long-slit spectra taken at
multiple position angles to estimate an overall rotational orientation and
velocity for the system. Our full spectroscopic coverage across the face
of the system allows us to take this further and infer the overall
geometry of the interaction, as well as resolving the gaseous kinematics
on smaller angular scales.

\begin{figure*}
\centering
\subfigure[]{\includegraphics[scale=0.50]{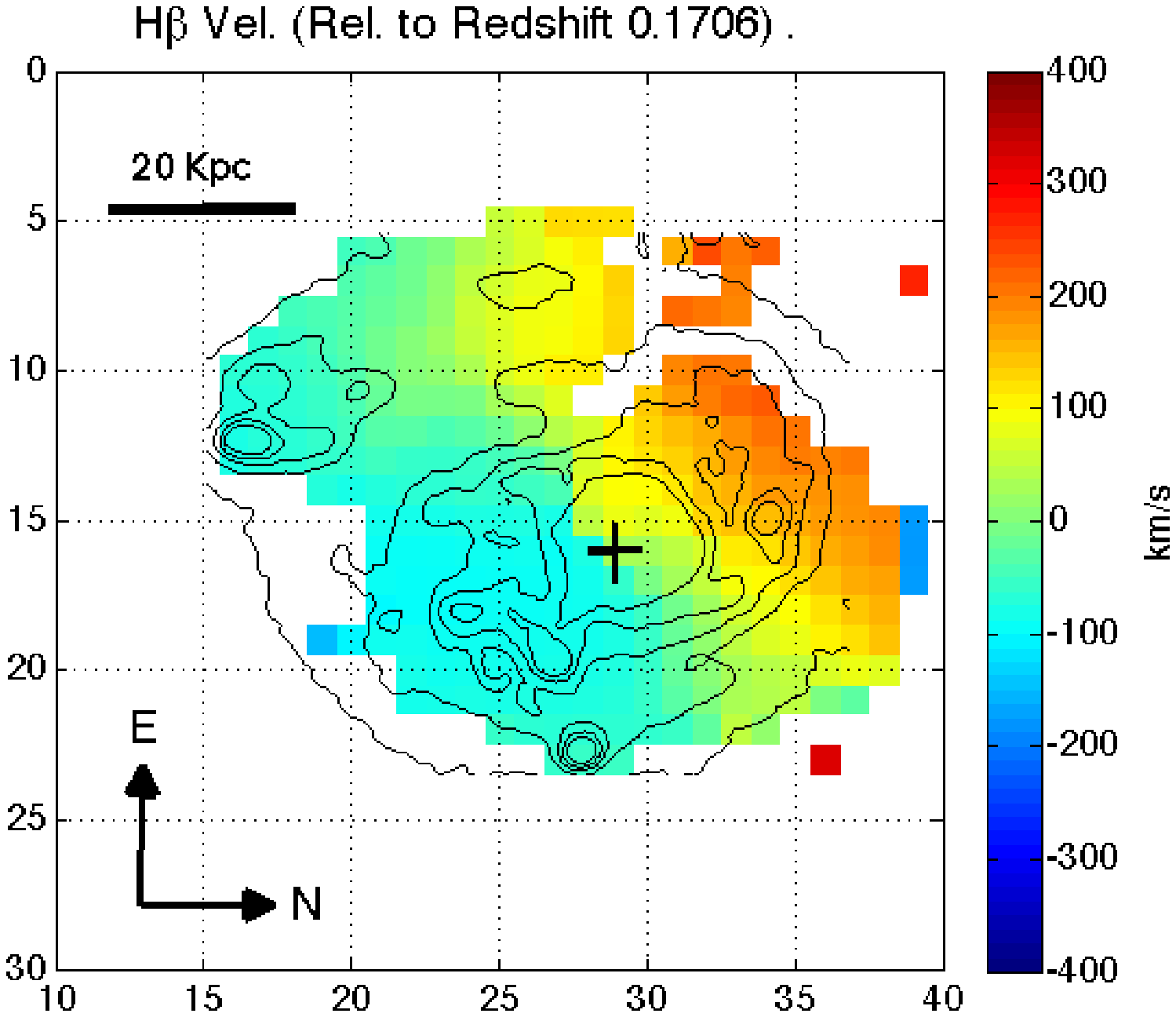}}
\hspace{0cm}
\subfigure[]{\includegraphics[scale=0.52]{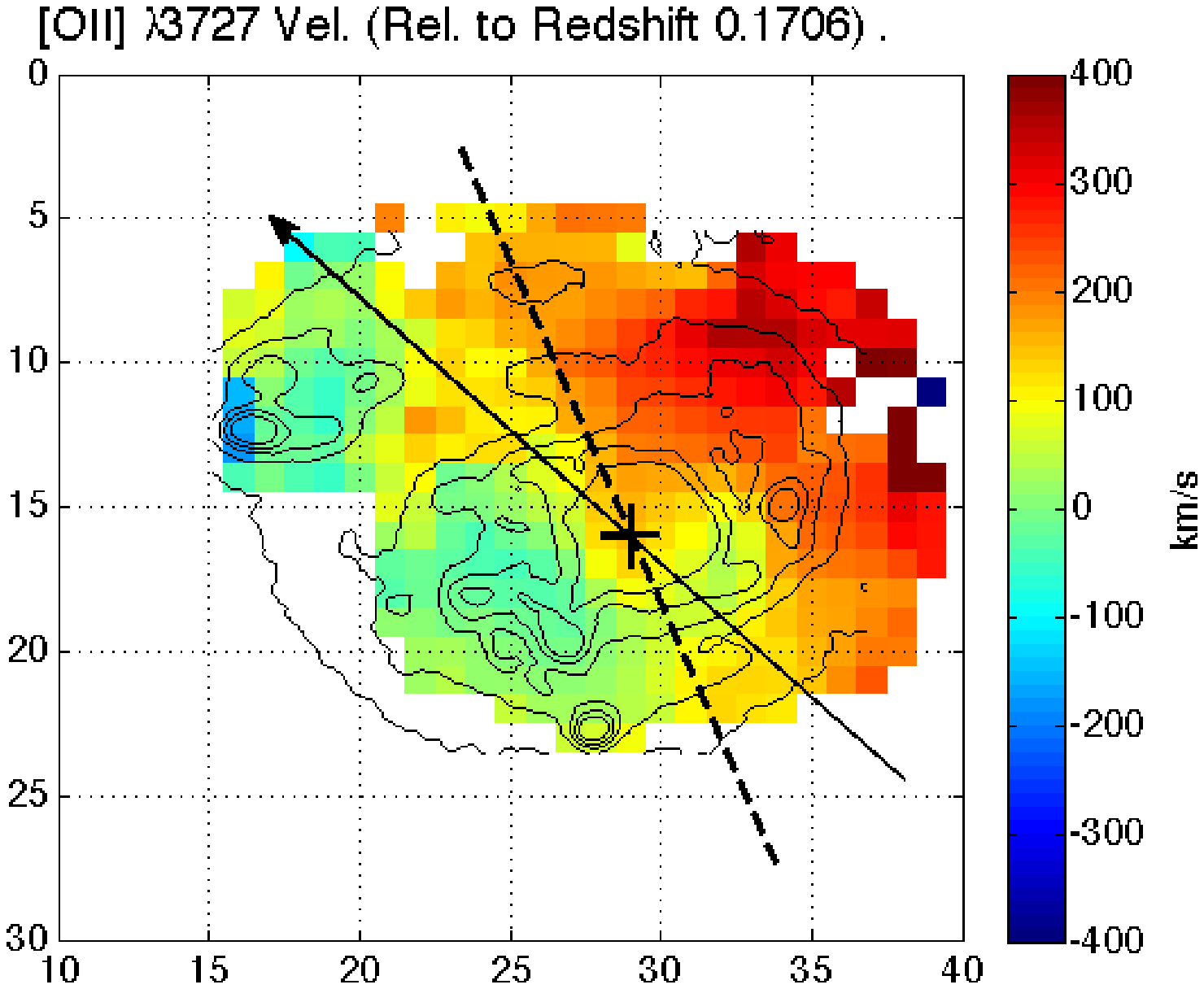}}
\hspace{0cm}
\subfigure[]{\includegraphics[scale=0.51]{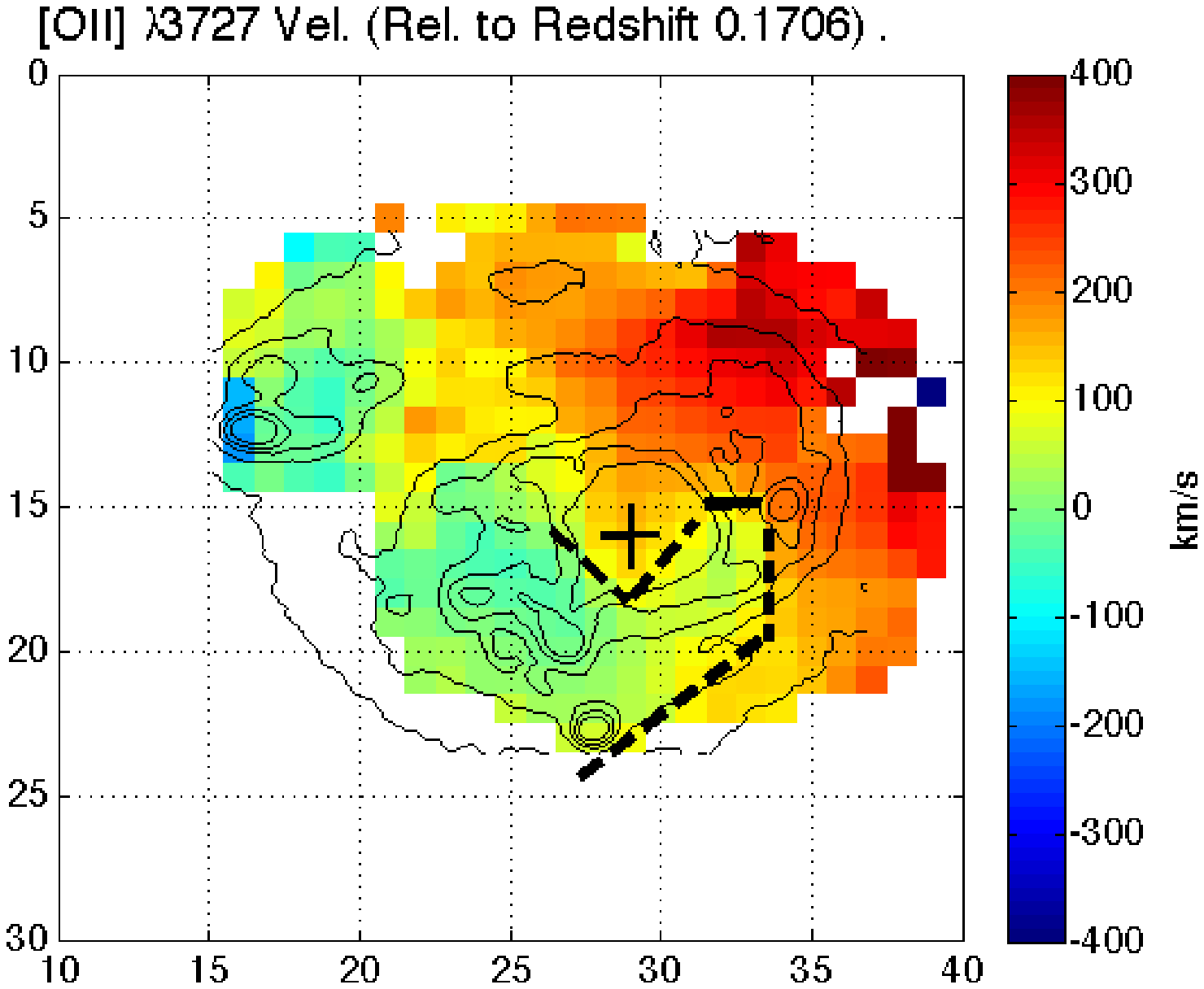}}
\hspace{0cm}
\subfigure[]{\includegraphics[scale=0.55]{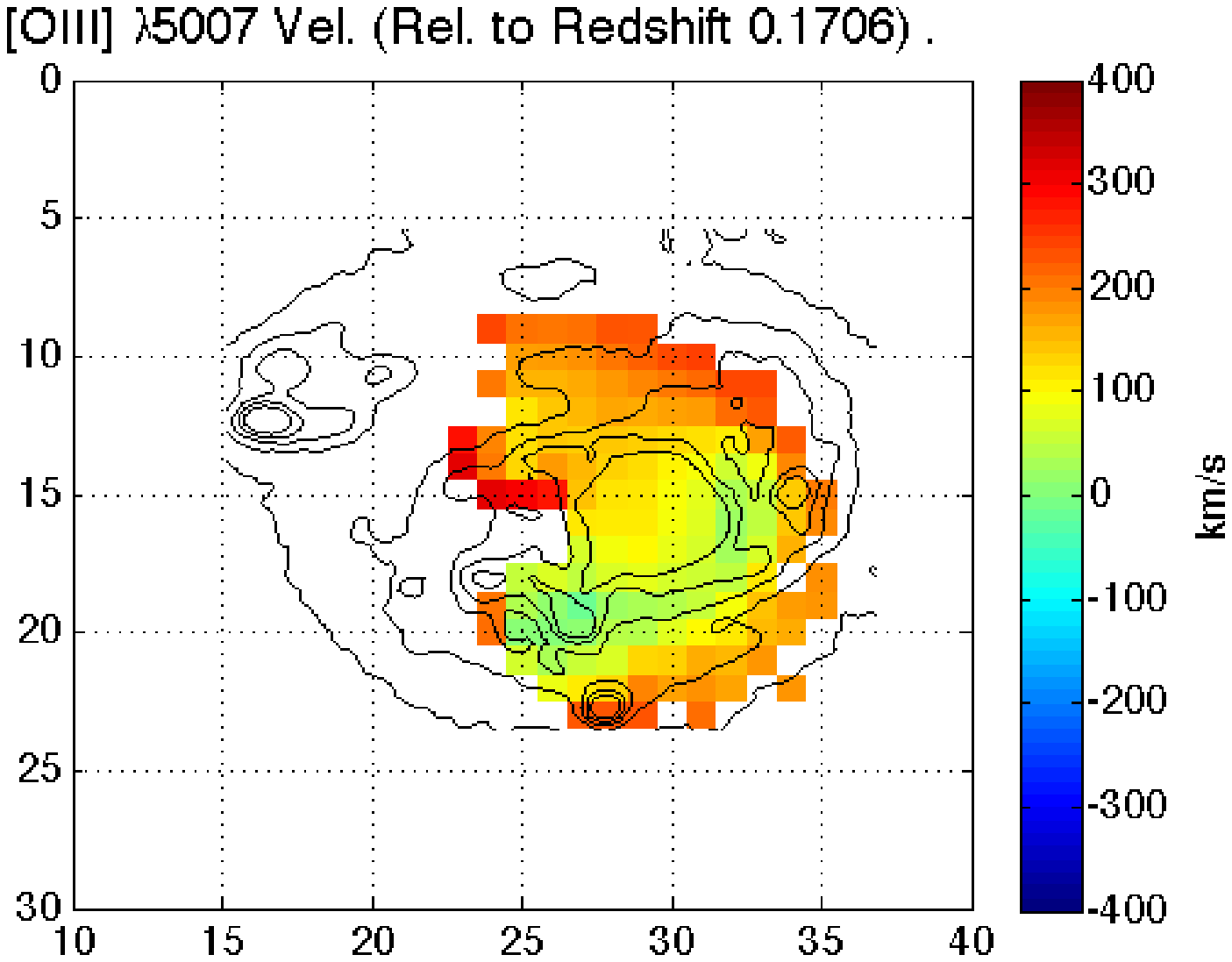}}
 \caption{\label{VelMaps} Colour-coded emission-line velocities for the
system, relative to $z=0.1706$, overlaid with brightness contours from the
$V$-band optical image. The axes refer to IFU spaxels, and a black
cross denotes the position of the galactic nucleus. a) H$\beta$, showing a
prominent rotation axis extending NW to SE through the system, and a
smooth rotation profile. b) [O\,II]$\lambda$3727, with the projected
rotation axis calculated from our IFU data using H$\beta$ emission
superimposed as a solid line (PA = $137^{\circ}$);  the rotation axis
determined by Safouris et al.\ (2003) is shown as a dotted line (PA =
$118^{\circ}$). c) [O\,II]$\lambda$3727, showing the blueshifted region
described in the text, delimited with dotted lines. d) the
[O\,III]$\lambda$5007, also showing a blue-shifted feature but no clear
rotation.}
 \end{figure*}

The reference frame adopted for the kinematic analysis was the redshift of
the system ($z=0.1706$; Simpson et al.\ 1993).  We computed the
relative velocity of the ionized gas in each spaxel, adjusted to the
heliocentric frame, and then created two-dimensional maps of the
line-of-sight velocity as a function of spatial position across the system
(Figure~\ref{VelMaps}), using the usual colour conventions for recession
and approach.  It is, of course, also possible to define the reference
velocity for each emission map in terms of the redshift of that emission
line at the galactic nucleus. In doing so, it is immediately clear that
there are significant differences between the two reference velocities.
For a discussion of the difference between the systemic and nuclear
emission-line redshifts, see Johnston et al.\ (2010).

The velocity maps resulting from our data show significant structure in
each of the major emission lines in our wavelength range. Furthermore,
each line shows a different velocity distribution, in some cases subtly
different, and in others markedly so. The H$\beta$ velocity distribution
shows a clearly defined rotational axis, with blueshifted regions to the
southwest of the nucleus and redshifted regions to the northeast. Using
the H$\beta$ data, we measure the position angle of the projected rotation
axis to be $137^{\circ}\pm5^{\circ}$ (see Fig. \ref{VelMaps}d). This
differs slightly from the PA obtained by Safouris et al. (2003), who
measured a PA of $118^{\circ}\pm10^{\circ}$ with a more limited data set
consisting of long-slit spectra. The apparent rotation is very smooth
across the system, characteristic of a uniformly rotating disk or sphere
of material. The largest relative velocity difference that we measure is
$\sim$280 km\,s$^{-1}$. The fact that we see significant rotation through
the entire system indicates that the companion is not orbiting in the
plane of the sky, as it appears in optical images, but the rotation axis
is significantly inclined to the line of sight.

 In comparison with H$\beta$, the oxygen lines show greater complexity in
their velocity structure. In [O\,II] we still see the same overall
rotation as in H$\beta$ but the nuclear region shows more complexity. The
most obvious deviation from the smooth H$\beta$ profile is the locally
blueshifted region of gas wrapping around the host galaxy nucleus (see
Fig.\ \ref{VelMaps}d), which we also observe in the [O\,III] emission.  
This feature appears to emerge out of the knotty region to the SW of the
nucleus, skirting the nuclear region to the W and NW. We speculate that
this structure may be associated with a large-scale flow of gas wrapping
around from behind the nuclear region, moving up towards us on an orbital
trajectory taking it around the nuclear region of the system.
Alternatively, we may be seeing a region of local kinematic complexity in
the gas, such as nuclear outflows, superimposed on the smooth underlying
rotation profile. We note that the single Gaussians employed to fit these
regions still provided an excellent fit to the emission lines, despite
this apparent outflow structure.  IFU data with higher spatial resolution
will be used to investigate this in more detail.

 We now digress briefly to consider how the kinematic results may
affect our analysis in section \ref{ionization}. The fact that the
different ionized species show different velocity structure implies that
the emission originates from different spatial regions in the host system.
We have therefore been careful to keep the interpretation of our
line-ratio diagnostics in section~\ref{ionization} rather general.
However, the sample of objects used to define the classification scheme in
Lamareille (2010) might potentially be expected to possess similar
ionization structure to MRC B1221$-$423. Given the scheme's firm basis in
empirical observation, valid over a diverse sample of objects, we believe
that our general conclusions are justified.  We now go on to discuss the
kinematics described above in the context of plausible interaction
geometries for this system.\\

\begin{figure*}\centering
\subfigure[]{\includegraphics[scale=0.40]{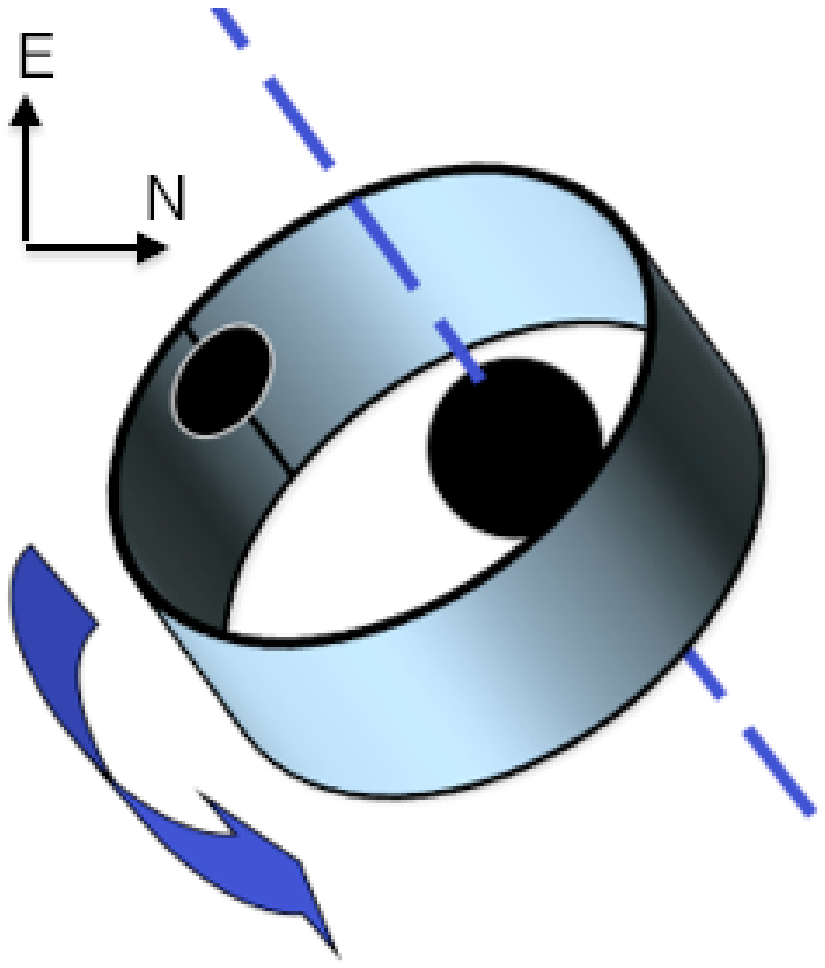}}
\hspace{0cm}
\subfigure[]{\includegraphics[scale=0.40]{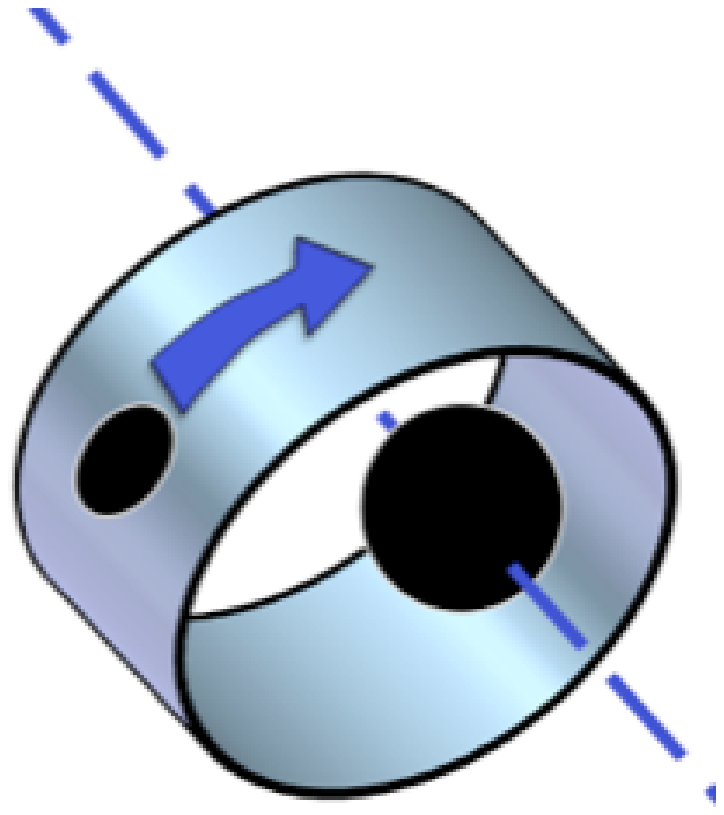}}
\hspace{0cm}
\subfigure[]{\includegraphics[scale=0.40]{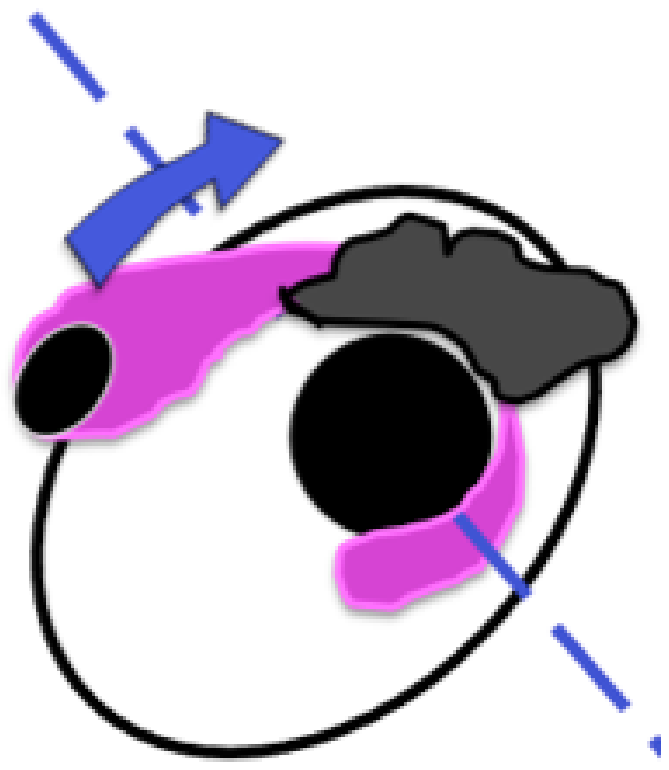}}
\hspace{0cm}
\subfigure[]{\includegraphics[scale=0.40]{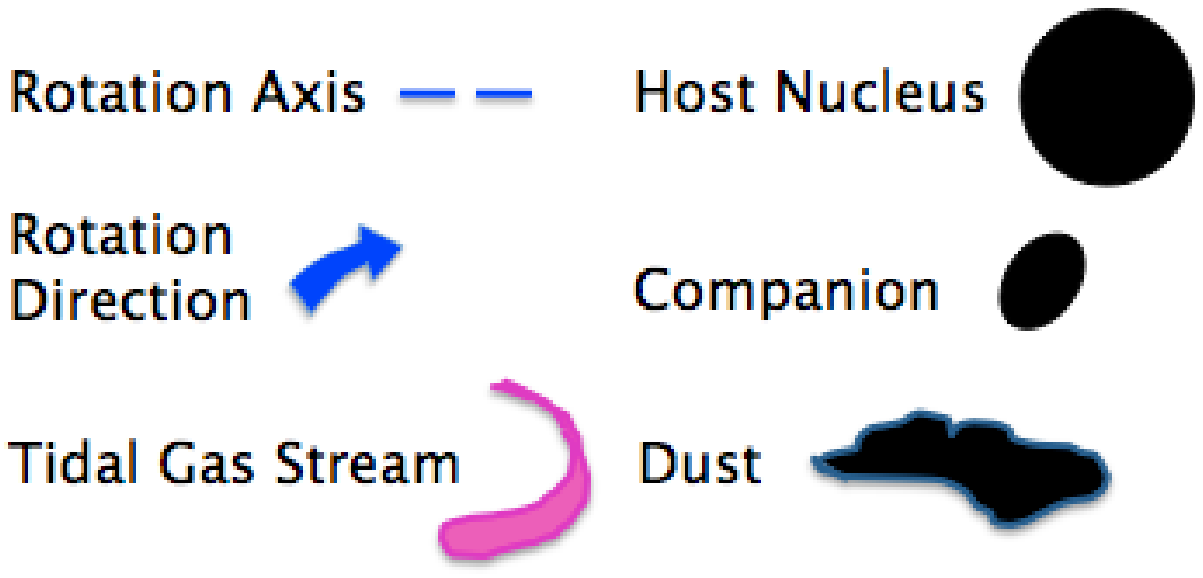}}
 \caption{\label{Models} Two toy models for the orbital geometry of the
MRC B1221$-$423 host and companion, inferred from the emission line
velocity data. a) \& b) show the two geometries for the companion's orbit
around the host allowed by the data, with the silver 'bands' in place to
make the orbital geometry clear. If the geometry depicted in b) is
correct, then c) shows a possible explanation for the \lq disappearing'
tidal tail described in section \ref{Distribution}, as well as the
blueshifted feature seen wrapping around the nucleus in the oxygen lines
(see figure \ref{VelMaps}d). In c), both of these features are generated
by a gaseous tidal flow from the companion into the host.} \end{figure*}

\subsubsection{Toy Models for the Interaction Geometry}
\label{ModelGeometries}

The line-of-sight velocity data allow two possible rotational senses for
the MRC B1221$-$423 system, and we now present toy models incorporating
each. These models aim to describe qualitatively the most significant
features observed in the kinematics and distribution of the gas. The first
model is presented in Figure~\ref{Models}a. It shows the host nucleus, and
the companion galaxy in orbit around it. The rotation axis extends through
the host nucleus where most of the mass in the system is contained,
running from the northwest to the southeast. In this scenario, the
companion is moving towards us and down to the southwest, after having
orbited around behind the system from our frame of reference. Referring
back to Figure \ref{ContourFluxMaps}a,b, the H$\beta$ tidal tail would
then arise as a result of being pulled out of the companion or host at or
near their point of closest approach. The knotty star-formation regions
could result from gas fed into the system or otherwise disturbed as the
companion has moved around on its orbit.

 There is a second viable alternative, broadly consistent with the
data, that we could not rule out (Figure~\ref{Models}b,c). In this
scenario, the companion is also moving towards us at present, but its
orbit is directed around in front of the system towards the northeast.
This picture is consistent with the velocity data, but possesses the
additional advantage of being able to explain the observation that the
tidal tail of H$\beta$ emission emanating from the companion disappears to
the east of the host nucleus (see Figure~\ref{ContourFluxMaps}b). In this
model, the orbit of the tidal gas stream takes it out behind the host
galaxy system, where it is possible that intervening dust could have
blocked the emission along our line of sight. Hence we see the stream
disappear from sight as it moves behind the galaxy, but then reappear as
it moves back up towards us on its trajectory around the nuclear region.
Such a gas flow might then be responsible for the blue-shifted region seen
in the oxygen lines and depicted in Figure~\ref{VelMaps}b, c \& d, and may
represent tidal gaseous accretion from the companion. Computational
modelling of such interactions (e.g. Mihos \& Hernquist 1996) shows that,
ultimately, the tidal debris can then sink into the core of the host to
fuel the AGN activity.

 These toy models can account for the broad kinematic behaviour of the
system. However, the kinematics rapidly increase in complexity as we probe
smaller scales and regions towards the nucleus, and our simple models are
clearly inadequate. Characterising the emission lines with multi-component
Gaussians shows that there are significant asymmetries in the line
profiles in the nuclear region, implying complex flows of gas driven by
high-energy phenomena. The complexity of these lines will require detailed
dynamical modelling and is beyond the scope of this paper.

\subsection{Modelling the Stellar Continuum}
\label{Continuum}

 Tidal disturbances caused by interactions between galaxies have often been
linked to episodes of widespread enhanced star formation (e.g., Sanders et
al. 1988, Mihos \& Hernquist 1996, Sanders \& Mirabel 1996, Veilleux
2001). Thus, the ages of stellar populations have the potential to shed
light on the interaction history of galaxies and, in turn, on the sequence
of events leading to merger-induced AGN triggering. To investigate the
history of the MRC B1221$-$423 interaction, we have modelled the ages of
the stellar populations in the system using combinations of Synthesised
Stellar Populations (SSPs).

\subsubsection{The Fitting Procedure}
\label{Fitting}

We used published isochrone spectral synthesis data from Bruzual \&
Charlot (2003) for our SSP models. Each SSP represents a synthesised
spectral energy distribution from an idealised population of stars that
formed together in an instantaneous burst of star formation. We assumed a
Salpeter initial mass function (Salpeter 1955) for the models, with masses
ranging between 0.1--100\,M$_\odot$, and solar metallicity. A total of ten
distinct SSPs were used in the fitting procedure, with evolved ages of 5,
25, 100, 290, 640, 900, 1400, 2500, 5000 and 11000\,Myr.

We constructed synthesised populations from these basis SSPs to model our
system by forming linear combinations of the old 11\,Gyr SSP with each of
the nine younger SSPs, in varying proportions. The result was an array of
model populations, with each column corresponding to a particular
combination of the 11\,Gyr population with one of the younger populations,
and each row corresponding to the relative proportion of the younger
population, increasing from 0\% to 100\% in 10\% increments. This allowed
us to model the base population of the galaxy ($\sim$11\,Gyr population)
plus one major starburst of a given age across the system. Sparse temporal
coverage at the older end of the SSP spectrum is justified on the basis
that populations of this age will principally contain low mass stars that
evolve only slowly with time.

We binned our datacube into ten distinct spatial regions to increase the
signal-to-noise ratio of the spectra for the fitting procedure. The
spatial extent of the binning regions was determined by features
discernible in optical images, such as the nucleus or star-formation
complexes (Figure~\ref{fig:BinRegionsa}). This was done to maximise
discrimination between different populations; distinct optical features in
the system are more likely to possess similar star-formation histories. As
described in section~\ref{ionization}, the spectra were corrected for
reddening intrinsic to the system using the Calzetti formulation. In the
fitting procedure employed here, however, the estimates derived by
Johnston et al.\ (2010) were used as a starting point for determining the
reddening parameter $A_V$.

We normalised and fitted each of the two-component SSP models in our array
to each of our spatially binned spectra, and then identified the best
reduced-$\chi^2$ fit(s) for each binned region. By careful comparison of
those models yielding the best $\tilde{\chi}^2$ fits, we were able to
identify population components that best characterised the binned spectra.
We verified that two-component models provided an adequate description of
the stellar populations in each binned region. Adding a third component
did not improve $\tilde{\chi}^2$, nor was the additional component ever
assigned more than 2\% of the total flux. Thus, the two-component model
was adequate to describe most of the spectra, with the exception of the
regions containing the host nucleus and companion, which are discussed
further below.

\subsubsection{Results of Synthesised Stellar Population Modelling}
\label{FittingResults}

The overlaid spectrum plots shown in Fig.\ \ref{Fits}(a),(c) and (e) are
representative of our fitting results, and demonstrate that excellent fits
can be obtained for the binned spectra over the entire wavelength range.
The models closely trace out features in the observed spectra and follow
both the detailed structure of specific spectral features, as well as the
more general shape of the spectra. An examination of the $\tilde{\chi}^2$
array plots for each region (Fig.\ \ref{Fits}(b),(d) and (f)) shows that
in most cases, a small range of ages is strongly preferred for a given
region, although the young component fraction is less well defined.  As
the principal aim of the modelling is to determine the approximate ages of
the underlying stellar populations, the fact that the proportions are less
well constrained than the ages does not significantly affect our
conclusions.

There is a degree of degeneracy between the proportion and age evident in
the $\tilde{\chi}^2$ plots, with the best $\tilde{\chi}^2$ values tending
to fall across diagonal bands, corresponding to different relative
proportions of younger and older stars. In cases where the
$\tilde{\chi}^2$ plot showed a range of potential age combinations with
similar $\tilde{\chi}^2$ values, the overlaid plots were manually
inspected to identify the best-fit model. Models providing a superior fit
to specific spectral features were accepted as more likely to characterise
a population than models that followed only the broad spectral shape.

\begin{figure*}
 \begin{minipage}[b]{0.5\textwidth}\
 \centering
 \begin{tabular}{ | l | l | l | l |}
 \hline
 Region No. & Region & Young Comp. & Young\\
 & Description & Age [Myr] & Comp. [\%]\\
 \hline
Region 1 & Companion & 640 & 80 \\
Region 2 & Tidal Tail & 25 & 30 \\
Region 3 & Tidal Tail & 25 & 30 \\
Region 4 & & 900 & 100 \\
Region 5 & & 640 & 60 \\
Region 6 & Host Nucleus & 900, 640, 100 & Varies \\
Region 7 & Knotty Region 1 & 640 & 100 \\
Region 8 & Knotty Region 2 & 100 & 40 \\
Region 9 & & 100 & 40 \\
Region 10 & & 640 & 40\\
 \hline   
\end{tabular}
\captionof{table}{Best-fit ages and luminosity contribution percentages 
of 
young stellar burst populations in MRC B1221$-$423,
associated with the full-coverage binning regions in Fig.\ \ref{fig:BinRegionsa}.}
\label{table1}
\end{minipage}%
\hspace*{6mm}
\begin{minipage}[b]{0.4\textwidth}
 \centering
 \includegraphics[width=5.5cm]{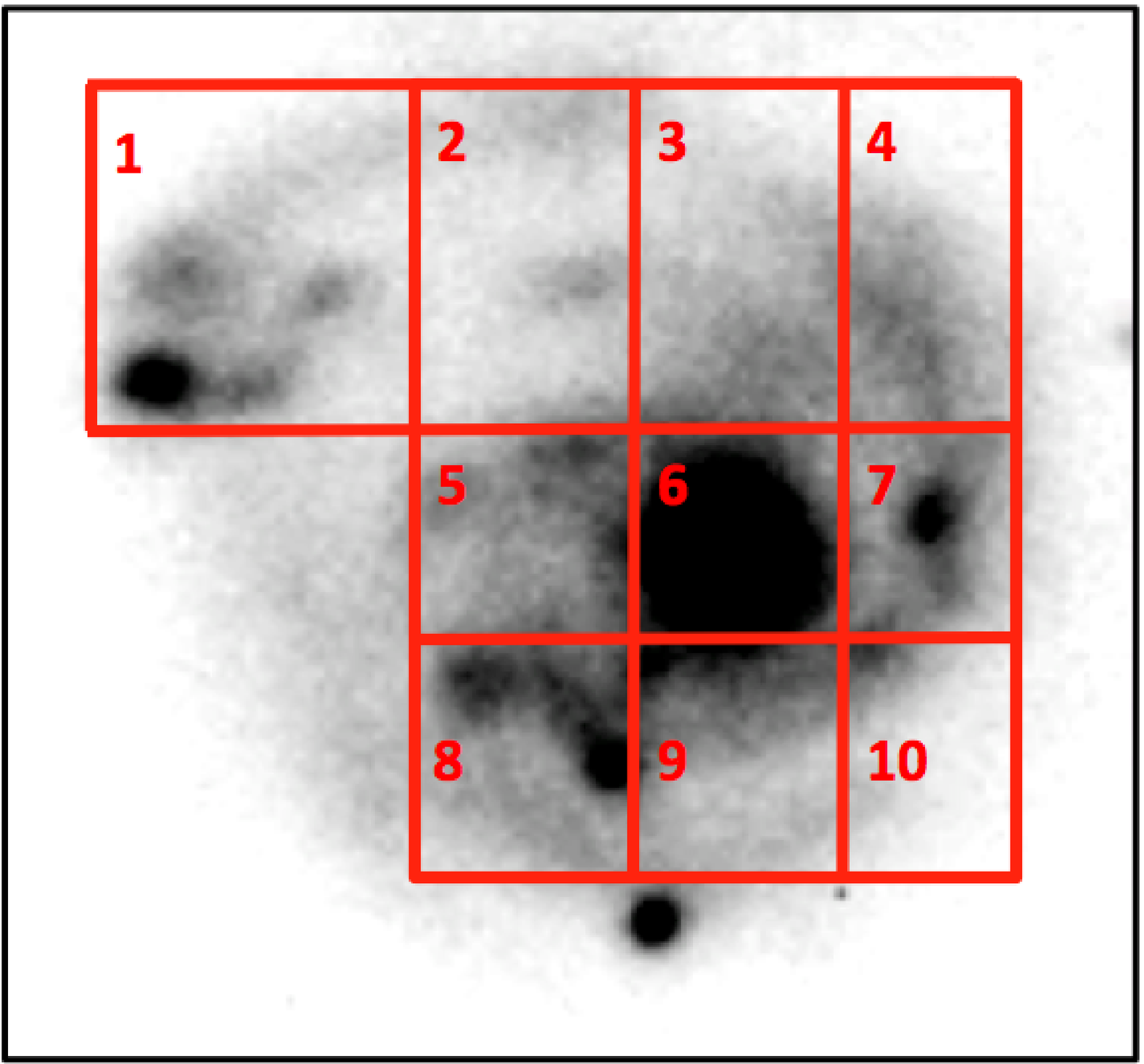}
 \captionof{figure}{Full-coverage spatial binning regions for isochrone
synthesis modelling, with best-fit parameters presented in Table
\ref{table1}}
 \label{fig:BinRegionsa}
 \end{minipage}
\end{figure*}

The results of our analysis are shown in Table \ref{table1}. They suggest
that each of the populations comprising the ten regions in Fig.\
\ref{fig:BinRegionsa} has undergone a significant star formation episode
in the last 1000 Myr, providing evidence for widespread disturbance to the
system during this period. The ages of the best-fit models are observed to
preferentially inhabit certain age bins, notably the bins corresponding to
the 640\,Myr and 100\,Myr starburst populations. Those falling into the
640\,Myr group include regions 1, 5, 7 and 10. The 100\,Myr group consists
of regions 6, 8 and 9.

\begin{figure*}
 \begin{minipage}[b]{0.5\textwidth}\
 \centering
 \begin{tabular}{ | l | l | l | l |}
 \hline
 Region No. & Region & Young Comp. & Young\\
 & Description & Age [Myr] & Comp. [\%]\\
 \hline
Region 1 & Tidal Tail (V-Band) & 25 & 30 \\
Region 2 & Tidal Tail (H$\beta$) & 25, 640 & 30, 2 \\
Region 3 & Knotty Reg. 1 & 640 & 100 \\
Region 4 & Knotty Reg. 2 & 100 & 40 \\
Region 5 & Companion & 640 & 100 \\
Region 6 & Extended [OIII] & 900 & 70 \\
Region 7 & Inner Nucleus & 640, 5 & 90, 2 \\
 \hline
\end{tabular}
 \captionof{table}{Best-fit ages and luminosity contributions of young
stellar burst populations in MRC B1221$-$423, associated with
morphological feature-specific binning regions in Fig.\ \ref{fig:BinRegionsb}}
 \label{table2}
 \end{minipage}%
\hspace*{6mm}
\begin{minipage}[b]{0.4\textwidth}
 \centering
 \includegraphics[width=5.5cm]{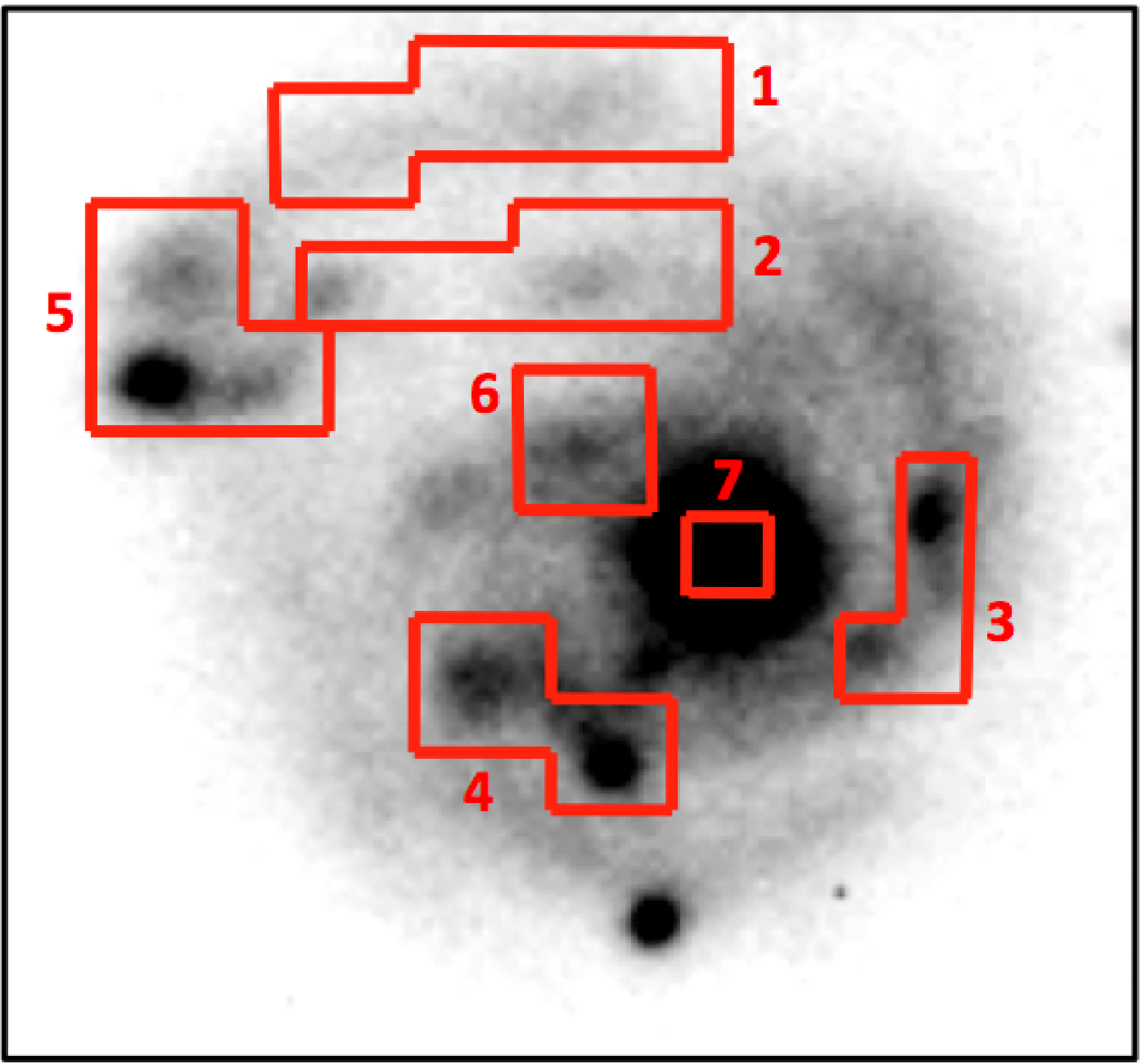}
 \captionof{figure}{Morphological feature-specific spatial binning regions
for isochrone synthesis modelling, with best-fit parameters presented in
Table \ref{table2}} \label{fig:BinRegionsb}
 \end{minipage}
\end{figure*}

 We repeated the fitting procedure with the more tightly defined regions
in Fig.~\ref{fig:BinRegionsb}. This allowed us to zoom in and examine the
specific morphological features identified in the V-band image. The
results for this second set of regions are given in Table\ \ref{table2}
and show that restricting the spatial extent of the fitting regions does
not significantly alter the age of the fit in cases where direct
comparison between the fit regions could be made.

We now discuss the models, with a particular focus on the interaction
history of this system. The tidal tail region of the companion (Fig.\
\ref{fig:BinRegionsb}, binning regions 1 and 2) is inherently dim, but the
binned spectra still provide adequate signal-to-noise levels. Young
populations clearly provide a superior fit to these regions, with a
substantial contribution from stars dated at $\sim$25\,Myr. This is not
unexpected, since star formation has often been observed in the gaseous
tidal tails of galaxies (e.g., Jarrett et al.\ 2006).

The host nucleus (Fig.\ \ref{Fits}(e)) and companion spectra show evidence
for more complex underlying stellar populations than the other regions.
Both show significant features that are not fully characterised by any
combination of the two-component models, though several combinations
provide close fits. More specifically, two-component models were unable to
fit all of the main spectral features. It is likely that these regions
have either seen multiple starburst episodes, or are currently undergoing
a significant amount of star formation. Both of these scenarios may apply,
and this is consistent with the picture of a prolonged period of
interaction between the two galaxies, as we would expect if the companion
galaxy has made multiple close passes.  See, for example, the behaviour
exhibited by the NGC 1512/1510 system (Koribalski \& L\'{o}pez-S\'{a}nchez
2009), and IRAS 08339+6517 (L\'{o}pez-S\'{a}nchez et al. 2006).

Another feature of note is that the stellar population in the extended
[O\,III] region described in section~\ref{Distribution} (see also Fig.\
\ref{fig:BinRegionsb}, region 6) shows no sign of a very young component.
Given the strength of the [O\,III] emission from this region, this
suggests that the ionization mechanism is not star formation, as discussed
in section~\ref{ionization}.

In section~\ref{Distribution}, it was suggested that one plausible cause
for the fact that we see separate tidal tails in H$\beta$ emission and the
V-band continuum is that these features could have been drawn out on
consecutive orbits, with the first orbit producing the V-band tail, which
subsequently formed stars and had the remainder of its gas stripped away.
The H$\beta$ tidal tail may then have been created on the most recent
encounter, with star formation just commencing. If correct, the population
ages of these two features should differ by an amount approximately equal
to the orbital period of the companion, estimated to be $\sim10^8$ years
(Johnston et al.\ 2010). However, we do not see any difference between the
fits in the two regions: both share a significant and very young
population of $\sim$25 Myr. The age of the stars in these features
indicate that these features were created during the most recent close
encounter, and that prodigious star formation is occurring in the tidal
wake of the companion.

Finally, a potentially interesting result peculiar to the fits derived for
region 6 in Fig.\ \ref{Fits}(d) is that by increasing the reddening
parameter $A_V$ from 2.2 to 3.2, the fitting routine will fit populations
as young as 5\,Myr. This possible young population only constitutes a very
small fraction of the SSP combination, but a component of this age would
be consistent with the findings of Johnston et al.\ (2010), who fit a
young 10\,Myr population to the broadband spectral energy distribution in
the host nucleus.

\begin{figure*}\centering
\subfigure[]{\includegraphics[scale=0.26]{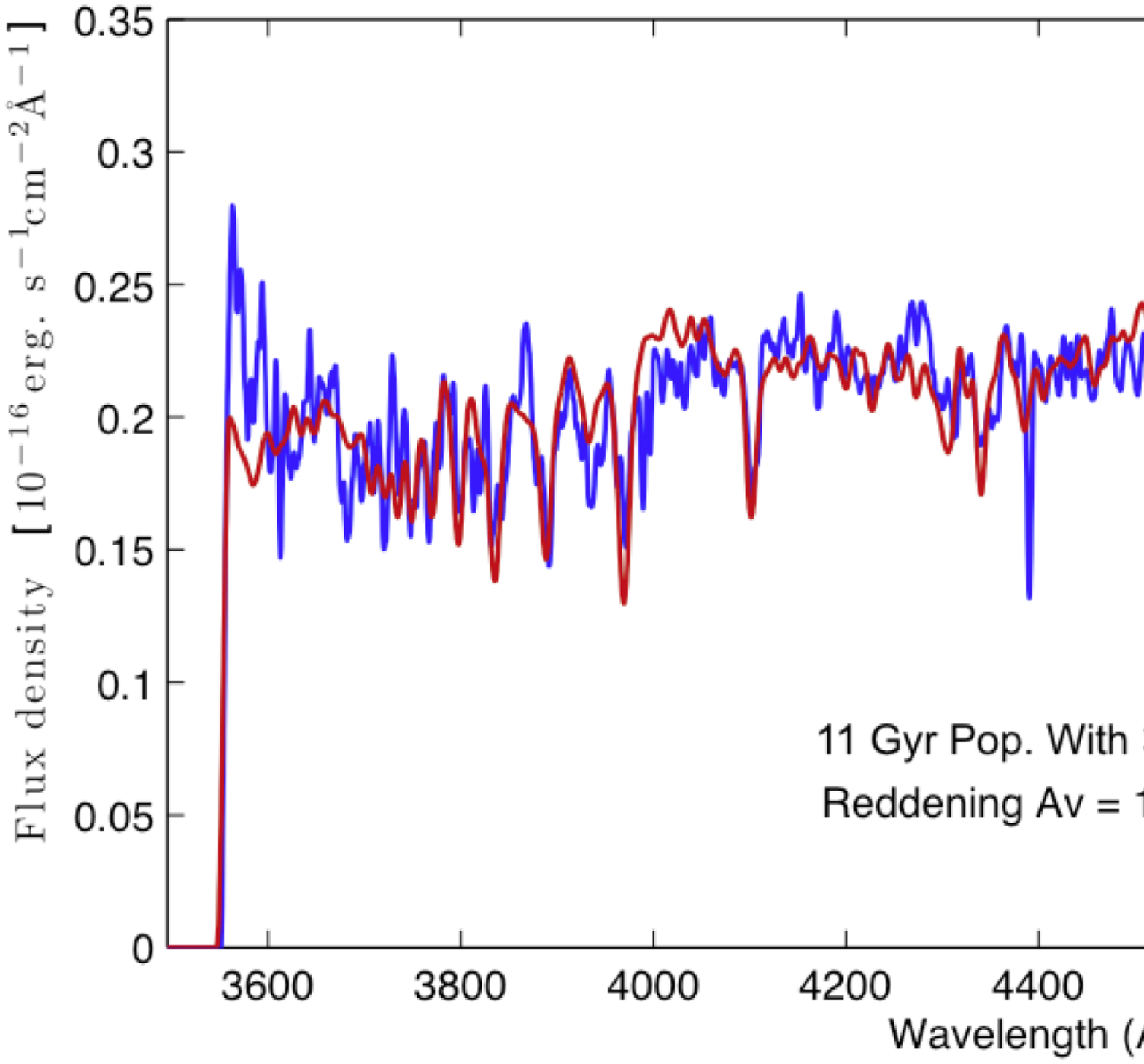}}
\hspace{0cm}
\subfigure[]{\includegraphics[scale=0.46]{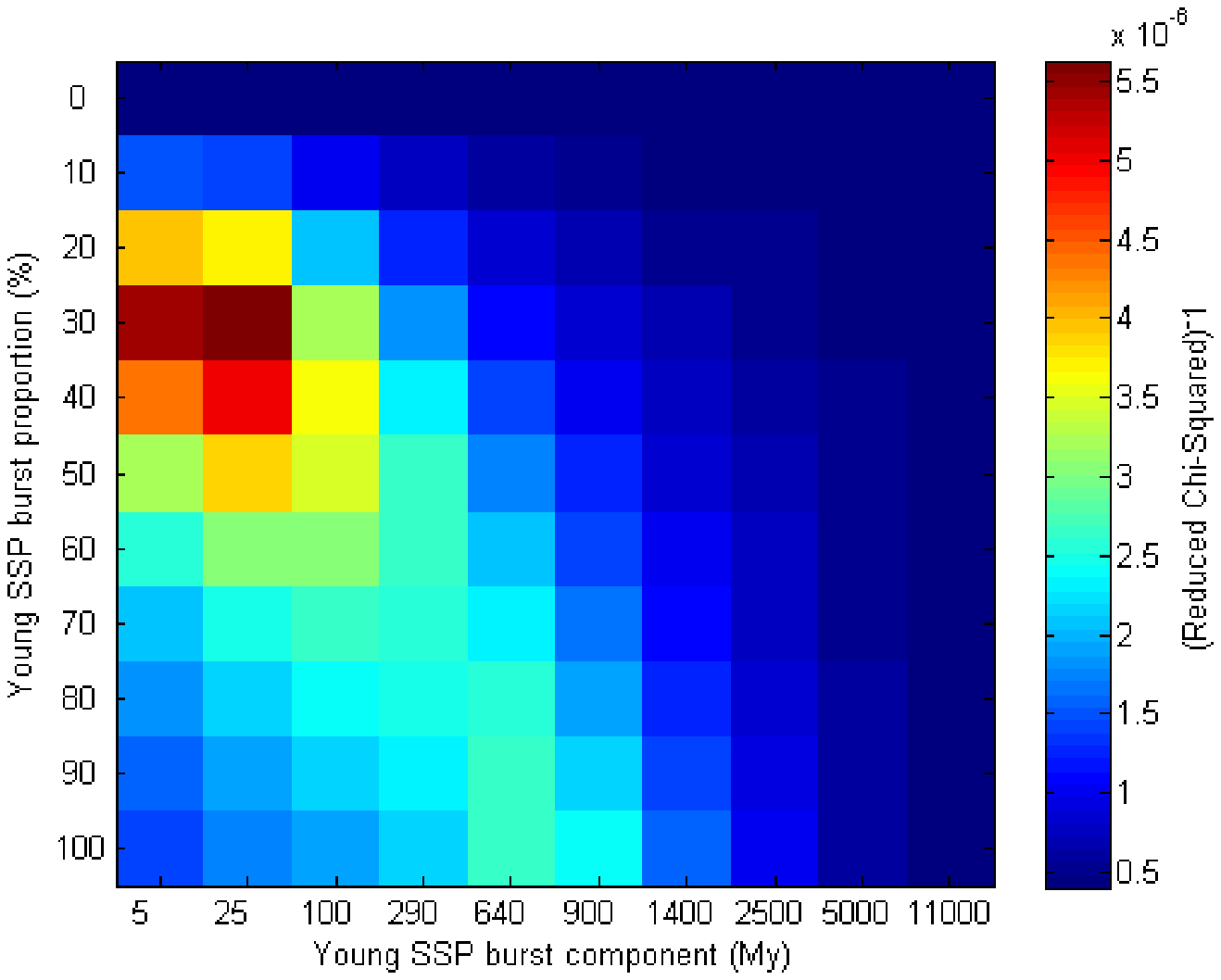}}
\hspace{0cm}
\subfigure[]{\includegraphics[scale=0.26]{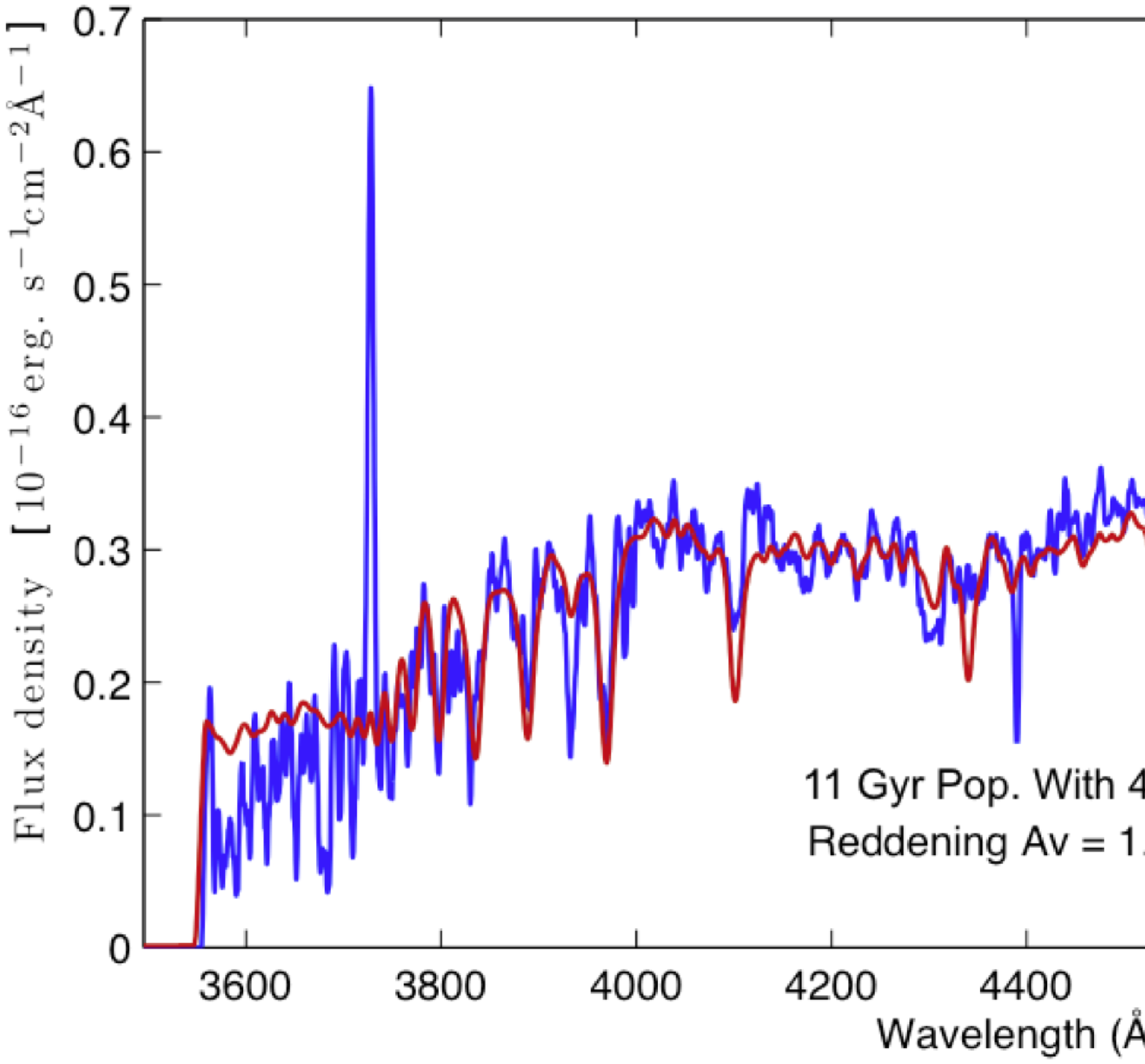}}
\hspace{0cm}
\subfigure[]{\includegraphics[scale=0.46]{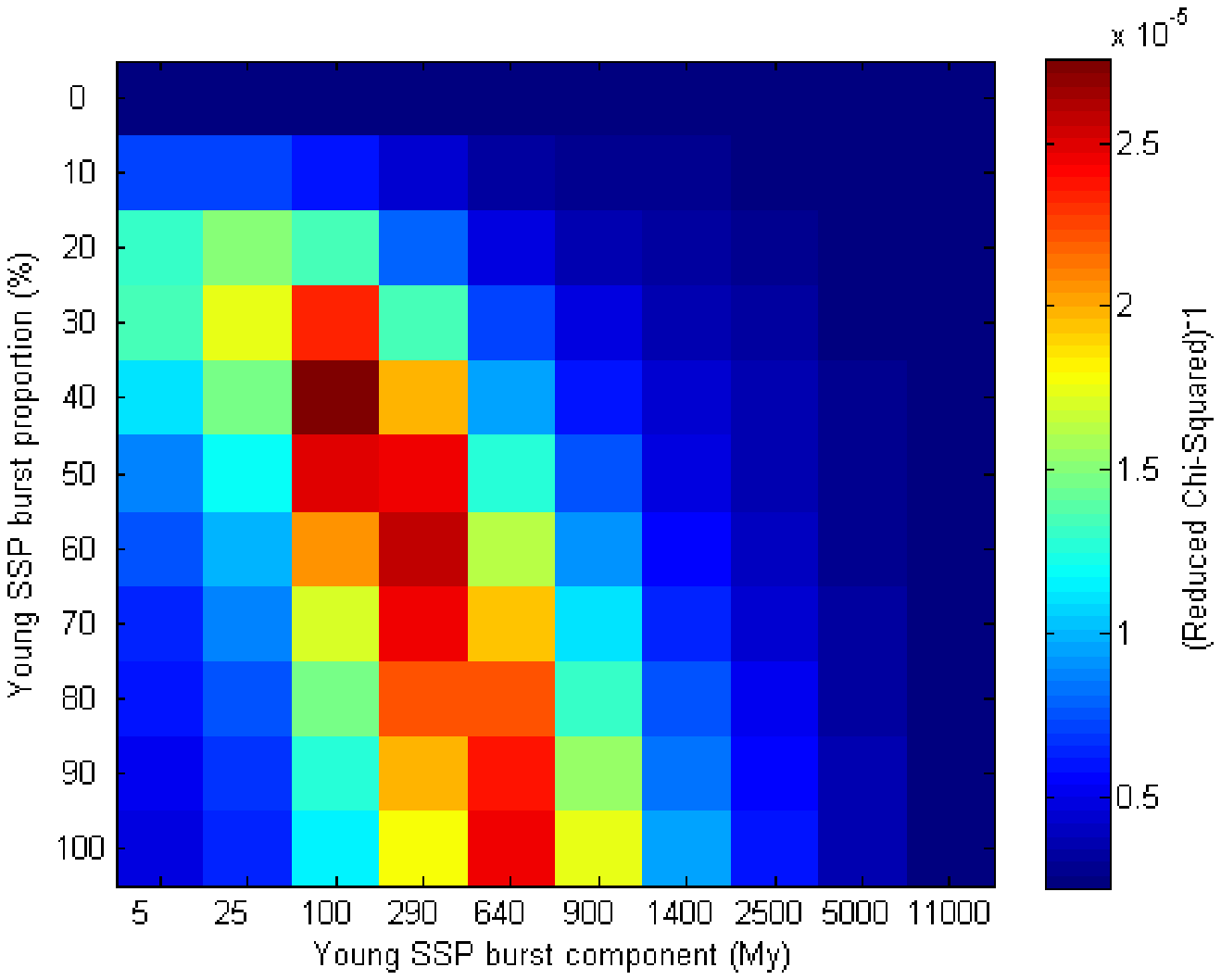}}
\hspace{0cm}
\subfigure[]{\includegraphics[scale=0.26]{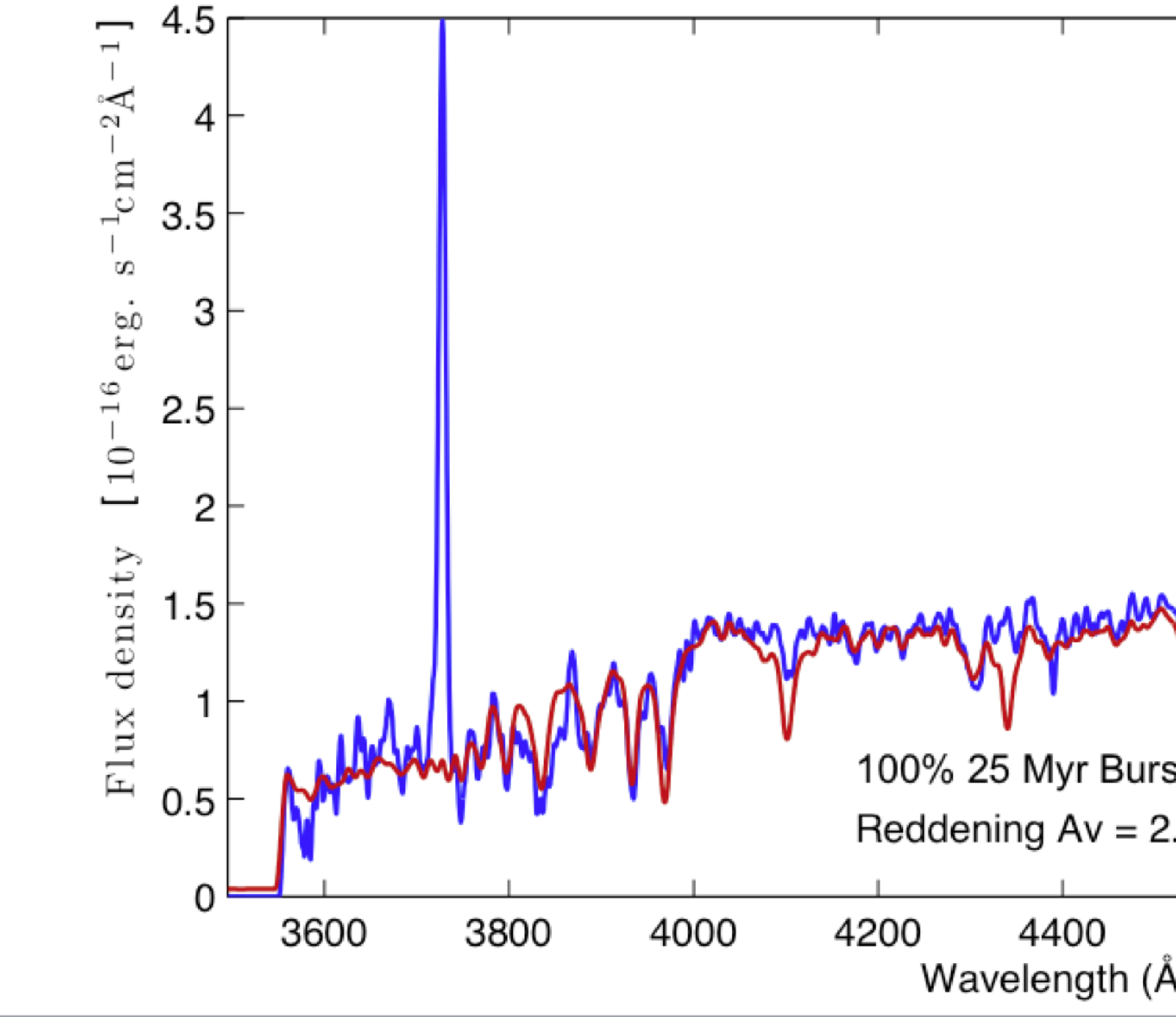}}
\hspace{0cm}
\subfigure[]{\includegraphics[scale=0.46]{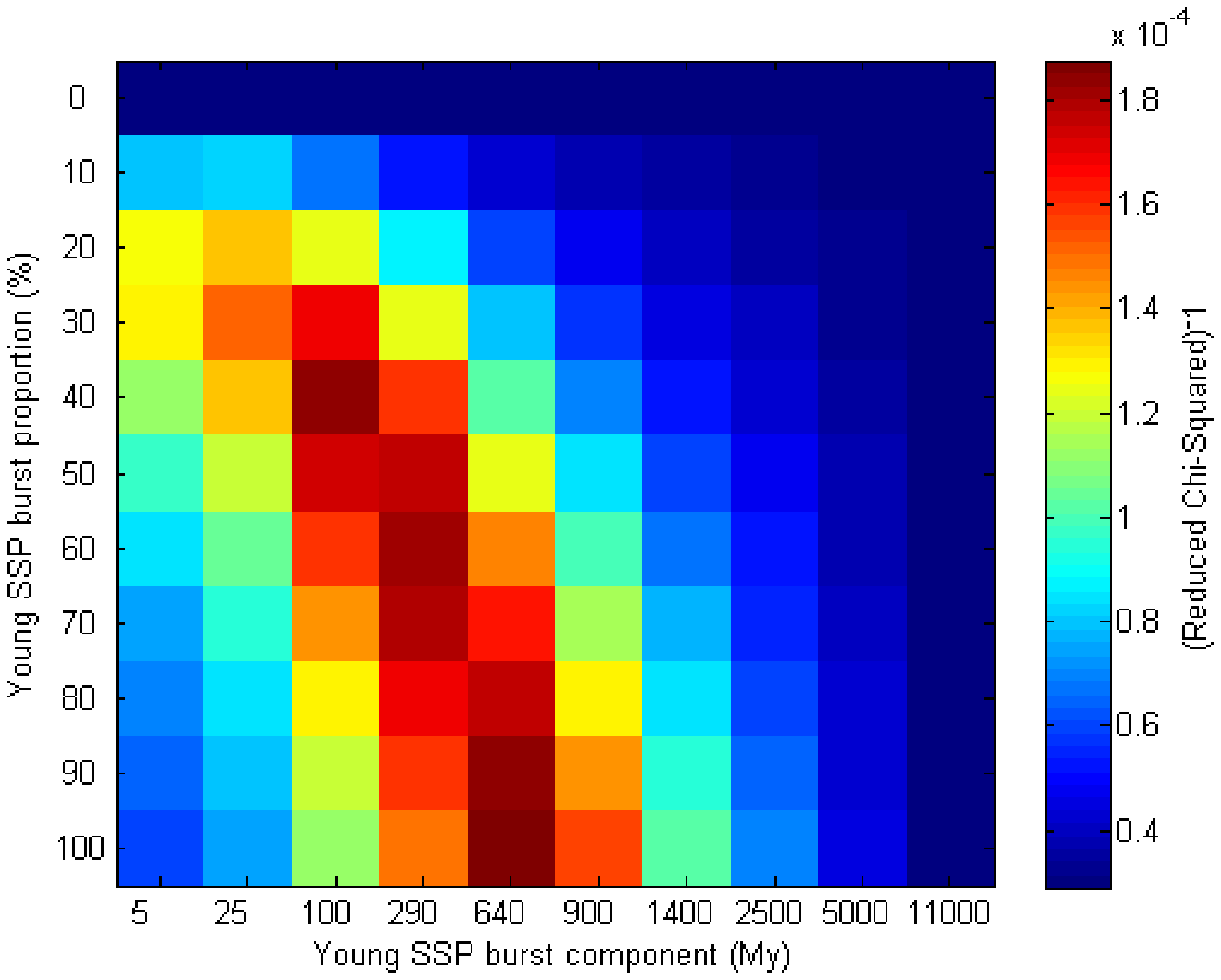}}
 \caption{\label{Fits} Model stellar population fits for selected binning
regions of the MRC B1221$-$423 system shown in Fig. \ref{fig:BinRegionsa},
with panels (a) \& (b) above calculated from binning region \#2 (tidal
tail), panels (c) \& (d) from binning region \#8 (knotty star-burst
region), and panels (e) \& (f) from binning region \#6 (host nucleus). The
blue trace represents the actual binned data (including emission features
not modelled by the SSPs), while the red trace shows the spectrum
generated by the best-fitting combination of SSP models. The colour plots
to the right show the inverse $\tilde{\chi}^2$ score for each SSP
combination tried (see section \ref{Fitting}). Higher values indicate
closer fits of the model to the data.} \end{figure*}

\section{Discussion and Conclusions}
\label{DiscussionConclusion}

We have presented an IFU study of MRC B1221$-$423, a system hosting a
young CSS source currently undergoing a minor merger. Our main findings
are:

\begin{itemize}
 \item strong emission from H$\beta$, [O\,III] and [O\,II] throughout the
system, with spatial distributions suggestive of both AGN-like processes
and extensive ongoing star formation
 \item diagnostic emission-line ratios indicating that AGN and
star-formation processes are associated with distinct morphological
features visible in broadband images
 \item smooth, rotation-like velocity profiles across the system in
H$\beta$, with more complex kinematic behaviour observed in [O\,II],
allowing us to develop toy models describing the interaction geometry of
the host and companion galaxies
 \item stellar population ages derived from SSP modelling indicating the
existence of distinct stellar populations of various ages occupying
spatially distinct regions in the system.
 \end{itemize}

The observed distribution of ionized H$\beta$ and [O\,II] emission
throughout the system is highly suggestive of widespread ongoing star
formation, while the [O\,III] emission appears to trace the presence of
higher excitation activity associated with the AGN. Although some of the 
[O\,III] emission may be coming from highly-ionised areas within star-formation
regions, some form of AGN-induced ionisation is favoured 
by the diagnostic line-ratio analysis  in
section~\ref{kinematics}. The line emission is spatially correlated with
the nuclear region of the host, the companion galaxy, and blue knotty
regions seen in the $V$-band optical image. These sites of active star
formation suggest recent, widespread disturbance and/or injections of gas
into the system. In particular, the association of this gas with what
appears to be a tidal tail linking the companion and host galaxy may
provide evidence of the sequence of events leading to the widespread star
formation. We put forward two plausible explanations for this apparent
tidal structure:

\begin{enumerate}
 \item that the companion has dragged the gas out as it has passed through
the outer envelope of the host, or
 \item that the gas is being tidally stripped from the companion by the 
host. 
\end{enumerate}

In either case, structures of this type might broadly be expected in light
of hierarchical merger simulations (e.g. Tissera et al. 2002). Modelling
also shows that much of this gas will eventually sink into the nuclear
regions of the host (e.g., Hernquist \& Mihos 1995). Thus, the ample
reserves of gas in this system clearly have the potential to fuel current
and future AGN activity.

The kinematic behaviour of the gas in the system was examined via the
radial velocities of each of the major lines within our spectral coverage.
On the broadest spatial scales, the data show a smooth rotation of the
system inclined to our line of sight. There is an ambiguity in the orbital
geometry of the system that we were unable to resolve. We have, however,
been able to use the data to arrive at two possible toy models for the
geometry of the interaction, illustrated in Figure~\ref{Models}. These
models provide a basic geometrical picture of the merger event, are able
to account for the basic morphological and kinematic structures, including
the tidal shells visible in V-band images, the structure of the tidal tail
joining companion and host, and the streams of gas apparent in our
kinematic analysis. The higher-excitation oxygen lines reveal a more
complex kinematic picture than that revealed in H$\beta$, particularly in
the nuclear regions. The nature of this complexity might perhaps shed
further light on the dynamics of the interaction, but this will require
detailed modelling beyond the scope of the current work. Overall, the
rotational kinematics are consistent with a minor merger interaction,
first posited for this system by Safouris et al. (2003), and built upon by
Johnston et al. (2005, 2010).

Our isochrone synthesis modelling in section~\ref{Continuum} shows
evidence for stellar populations of different ages inhabiting spatially
distinct regions of the host and companion. We find that population ages
throughout the system cluster about specific values, particularly 100 Myr
and 640 Myr, albeit with coarse temporal resolution. There is some
evidence hinting at a young population of 5\,Myr in the nuclear region of
the host, but higher spatial resolution will be required to confirm this.
The ages derived in our study differ from some of those obtained by
Johnston et al. (2005, 2010). However, given the significant differences
in spectral and spatial coverage between the two techniques, this is
perhaps not surprising.

Metallicities derived across the MRC B1221$-$423 system are similar for
most regions (see Table \ \ref{intfluxtab}).  However, the fact that the
metallicity for the companion is marginally the lowest (notwithstanding
the 0.1--0.2 dex errors typical of the method used), followed by
that for the region containing the significant 100\,Myr-old stellar
population, is consistent with the ages and sequence of events we propose
below.

\begin{figure}
\begin{center}
\includegraphics[scale=0.6]{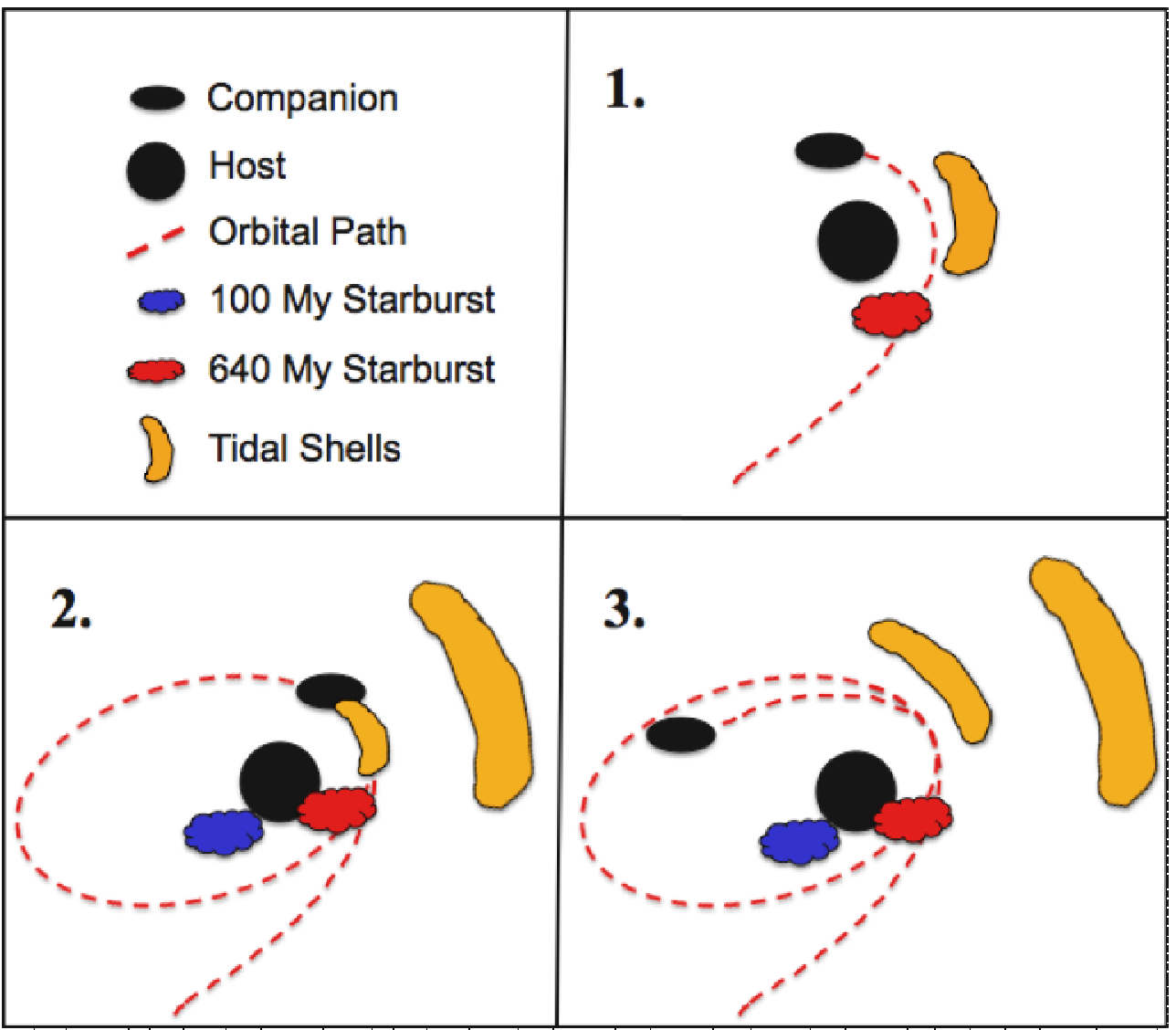}
\includegraphics[scale=0.215]{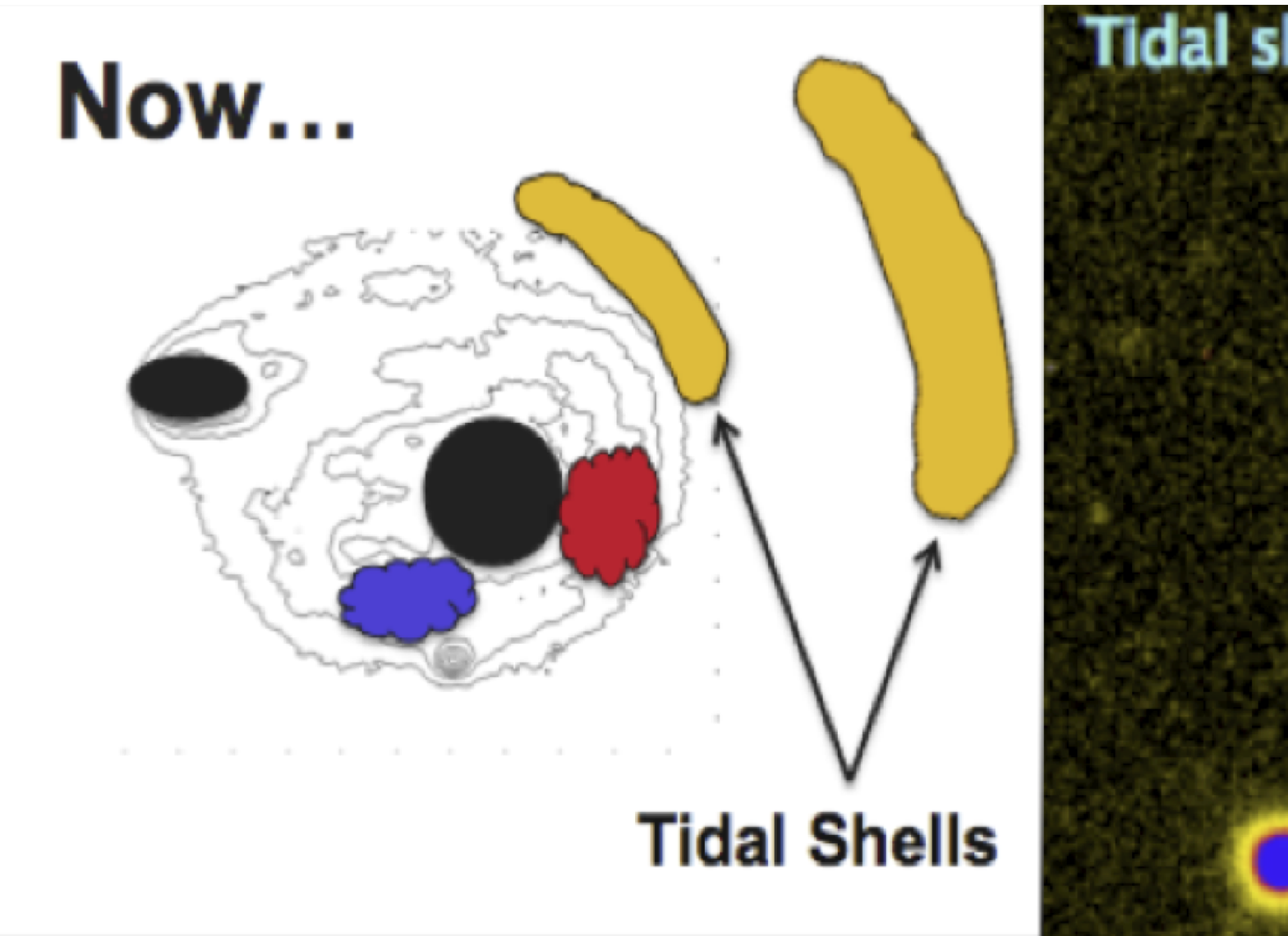}
 \caption{\label{SequenceEvents}Proposed sequence of events in the
triggering of MRC B1221$-$423, based on results described here, and by
Johnston et al.\ (2010), Johnston et al.\ (2005), and Safouris et al.\
(2003). The widespread ongoing star formation, the prominent starburst
populations at 640\,Myr and 100\,Myr, the two shells of tidally disrupted
material visible in broadband images, the tidal tail being pulled out by
the companion and the recently triggered CSS source are all consistent
with the sequence of events associated with the minor merger scenario
proposed in this figure.}  
 \end{center} 
 \end{figure}

Given the data accumulated for this system from photometry, spectroscopy
and radio imaging, we now present a possible timeline and sequence of
events leading from the initial interaction of the two galaxies through to
the present epoch. The approximate orbital period of the companion was
estimated by Johnston et al. (2010) to be $\sim10^8$ years. We now suggest
that this estimate may be too low. If the companion was initially captured
into a highly eccentric orbit of period $\sim$3-5$\times10^8$ years,
bursts of intense star formation (identified by our stellar population
modelling) may have been triggered by consecutive strong interactions near
the periapses of the companion's orbit. On this basis, we propose the
following sequence of events, depicted schematically in Figure
\ref{SequenceEvents}:

1) The companion makes an initial close pass by the host galaxy,
interacting strongly with it. Gas is either stripped from the companion by the
host, or gas already within the host is dynamically disturbed. In either case, 
it subsequently begins forming stars---the 640 Myr-old starburst. Some 
gas also begins to sink down into the nuclear region of
the host. A shell of stars is thrown out of the host, the companion loses
orbital energy, and becomes trapped into an eccentric orbit.

2) The companion then moves out on its orbit to apogalacticon and then
back to perigalacticon once again, taking somewhere on the order of
$\sim$3$\times10^8$ years. The ensuing interaction triggers a second
major episode of gas stripping/star formation---the 100\,Myr-old
starburst---and throws out a second shell of stars. The tidal tail
visible in optical images begins to be pulled out of the host or companion
at this stage. Some of the stripped gas sinks to the core of the host
galaxy, fuelling the supermassive black hole and triggering the radio-loud
AGN.

3) We find the system in its current state about 100\,Myr after this
latest perigalacticon, and can compare it to the false-colour optical
image in Figure~\ref{SequenceEvents}. We can identify the two shells of
stars pulled out of the system, and stellar populations resulting from
star bursts corresponding to the periods of most intense interaction
approximately 100\,Myr and 640\,Myr ago. The CSS radio source has recently
been switched on, after the gas has had $\sim$100\,Myr to sink into the
nuclear regions, in approximate agreement with the timescales over which
such processes are thought to occur (e.g., Lin, Pringle \& Rees 1988).
Johnston et al. (2010) show that the CSS source is potentially undergoing
a dramatic interaction with a dense surrounding environment --- the gas
stripped from the companion that has been drawn towards the nucleus of the
host galaxy.

 Projecting into the future, the companion is expected to continue to
lose orbital energy and merge fully with the host.

\section*{Acknowledgments}

The work herein is based on observations made with ESO Telescopes at the
La Silla and Paranal Observatories. The data in this paper have
been reduced using VIPGI, developed by INAF Milano, in the framework of
the VIRMOS Consortium activities.\\

 The authors would like to thank the referee \'{A}ngel
L\'{o}pez-S\'{a}nchez for detailed and helpful comments that have
significantly improved the quality of this paper, and which have informed
ongoing work on this object.  We also thank Rob Sharp (Australian
Astronomical Observatory and Research School of Astronomy and
Astrophysics, ANU) for providing custom analysis software in support of
this work, and Brooke Steel for her assistance in the preparation of
figures.

\label{lastpage} 

\end{document}